\documentclass[aps,pra,reprint,twocolumn,superscriptaddress,longbibliography]{revtex4-2}

\usepackage{amsmath,amsfonts}
\usepackage{MnSymbol}
\usepackage{hyperref}
\hypersetup{
  colorlinks=true,
  linkcolor=blue,
  filecolor=blue,
  citecolor=blue,
  urlcolor=cyan,
}
\usepackage{xcolor}
\usepackage{braket}
\usepackage{graphicx}
\usepackage{siunitx}
\usepackage{orcidlink}
\usepackage{mathrsfs}

\begin{document}

\title{Interaction geometry and ground-state properties of sparse quantum lattice models}

\author{Alex Gunning\,\orcidlink{0009-0002-9028-3413}}
\affiliation{Department of Physics, Clarendon Laboratory, University of Oxford, Parks Road, Oxford OX1 3PU, United Kingdom}

\author{Sebastian Schmid}
\affiliation{Department of Physics and SUPA, University of Strathclyde, Glasgow G4 0NG, UK}

\author{Zhengxiao Liu}
\affiliation{Department of Physics, Clarendon Laboratory, University of Oxford, Parks Road, Oxford OX1 3PU, United Kingdom}

\author{Sridevi Kuriyattil\,\orcidlink{0000-0002-9813-4714}}
\affiliation{Department of Physics, Clarendon Laboratory, University of Oxford, Parks Road, Oxford OX1 3PU, United Kingdom}

\author{Aydin Deger\,\orcidlink{0000-0002-6351-4768}}
\affiliation{Department of Physics, Clarendon Laboratory, University of Oxford, Parks Road, Oxford OX1 3PU, United Kingdom}

\author{Andrew J. Daley\,\orcidlink{0000-0001-9005-7761}}
\affiliation{Department of Physics, Clarendon Laboratory, University of Oxford, Parks Road, Oxford OX1 3PU, United Kingdom}

\begin{abstract}
We investigate how interaction geometry shapes the low-energy phases of sparse tunable long-range quantum models. We focus on a class of graphs whose degree grows logarithmically with system size, and show how symmetry and frustration in graph connectivity can drive, suppress, and reshape ground-state phase transitions. The central examples are power-of-$p$ graphs, where even and odd values of $p$ exhibit qualitatively distinct behaviour: even-$p$ graphs inherit the rich phase structure of the power-of-two model, while odd-$p$ graphs are governed by geometric frustration. Fibonacci graphs provide a contrasting case, lacking the discrete self-similarity of the power-of-$p$ family but exhibiting a direct geometric mapping between the short- and long-range limits. Across our models, we find that phase structure and criticality are governed by the same effective-geometry principle, unifying our framework for experimentally motivated long-range quantum systems.
\end{abstract}

\maketitle

\section{Introduction}
Long-range interacting quantum systems are now routinely engineered in atomic, molecular, and optical platforms, enabling coupling graphs that extend well beyond the local, nearest-neighbor graphs that historically motivated most condensed-matter models \cite{britton_2012,Islam2013,zeiher_2017,vaidya_2018,hollerith_2022} . For lattice models, long-range interacting systems are commonly defined by an interaction strength that depends algebraically on distance, $J(r) \propto r^s$. Treating $s$ as a continuous parameter has enabled theoretical exploration of unconventional phase transitions and dynamical regimes \cite{Dutta_2001,Koffel_2012,Daley_2013,Pupillo_2014,Defenu_2017, bentsen_2019}. 

Cavity-QED platforms now effectively realize such long-range couplings with continuously tunable distance dependence, making it natural to view interaction range itself as a controllable resource \cite{Hung_2016,Bentsen_2019_sparse,qin_2019,Periwal_2021,Mivehvar_2021}. These platforms offer control not only over the decay profile but also over the inclusion of individual couplings. In particular, one can selectively remove interactions at chosen distances, turning a fully connected long-range model into a deliberately \textit{sparse} coupling graph \cite{bentsen_2019}. Complementary approaches to realizing sparse long-range graphs have also emerged in Rydberg tweezer platforms, where atom shuttling protocols can selectively generate interactions between chosen sites \cite{Bluvstein_2022,Graham2022,Bluvstein_2023,Evered2023,Xu_2024,Bernien2017}. More recently, a dual-species Rydberg architecture was proposed in which the effective geometry of the interaction graph is encoded directly into the physical geometry of the atomic array, allowing geometry-driven phase transitions to be implemented \cite{gunning_2025}. Together, these capabilities open the door to a broad class of exotic sparse interaction graphs that remain both experimentally motivated and theoretically rich.
 
This paper focuses on a realizable and broadly motivated class of sparse long-range graphs with logarithmically growing degree: in a system of $N$ sites, each site couples to only $\mathcal{O}(\log N)$ others. Two well-studied examples are the power-of-two graph \cite{bentsen_2019,Bentsen_2019_sparse,Hashizume_2022,Kuriyattil_2023,Kuriyattil_2025,gunning_2025}, in which sites connect when their separation along a 1D ring is a power of two, and the hypercube graph \cite{Hashizume_2021,Kuriyattil_2023,Kuriyattil_2025}.  These systems provide examples of how sparse interaction graphs can give rise to “all-to-all-like” behavior, including fast scrambling \cite{bentsen_2019,Bentsen_2019_sparse,Hashizume_2021}, metrologically useful entanglement \cite{Kuriyattil_2025}, rich dynamical and low-energy phases \cite{Kuriyattil_2023,gunning_2025}, and suppression of disorder-induced localisation \cite{singh2026}. Here, we move beyond these canonical graphs to investigate whether their distinctive properties persist across a broad class of sparse long-range graphs. In particular, we exploit graph connectivity to enforce (and break) frustration and symmetry, which shape the many-body behaviour and low-energy phases in both spin and fermionic settings.

To address this, we first generalize the power-of-two construction to a broader family of power-of-$p$ graphs. Varying $p$ changes the opportunities for geometric frustration and thus the resulting many-body physics, while preserving a discrete self-similarity in the connectivity: if a bond exists at distance $d$, then a bond exists at distance $pd$. We then deliberately step outside this self-similar class by studying the Fibonacci graph, where bonds occur between sites at Fibonacci separations. The Fibonacci construction breaks the rigid scaling symmetry of power-of-$p$ graphs and provides a complementary, purposefully less regular benchmark within the same sparsity class.

The paper proceeds as follows. We begin by choosing a graph from our family of sparse logarithmic graphs, defined in Section~\ref{sec:graphs}. This step is Hamiltonian-agnostic: our analysis follows a geometry-first approach, building on the effective-geometry perspective introduced in \cite{gunning_2025}, which predicts much of the low-energy structure directly from graph connectivity alone. 
\begin{figure*}
    \centering
    \includegraphics[width=0.8\linewidth]{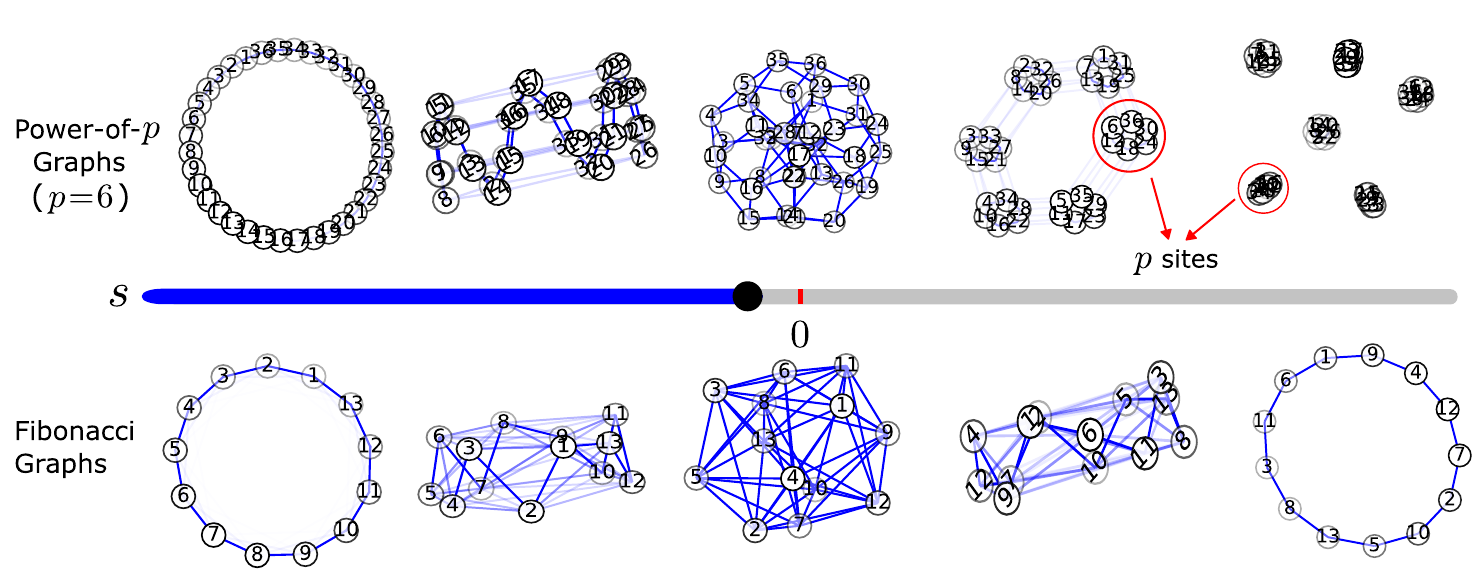}
    \caption{\textit{Slider schematic of the geometry underlying the sparse long-range models as a function of the tuning parameter $s$.} For the power-of-$p$ construction (top), varying $s$ continuously deforms the effective interaction geometry between the $N=p^l$ sites, interpolating between a ring-like structure, intermediate “tambourine”-like connectivity, a sparse all-to-all configuration, and finally a set of disconnected loops of size $p$. Here $p=6,\; l=2,\; N=36$. The Fibonacci construction (bottom), defined on $N$ sites with $N$ a Fibonacci number, exhibits a symmetric evolution about $s=0$, transitioning from a ring-like geometry through increasingly connected structures to a maximally connected configuration and back again. Here $N=13.$}
    \label{fig:effective geometries}
\end{figure*}
After establishing the relevant geometric structure for each graph family, we introduce in Section~\ref{sec:our models} two concrete models defined on these graphs: an antiferromagnetic transverse-field Ising model and a spinless fermionic hopping model. The latter provides an exactly solvable point of comparison, offering detailed insight into the ground-state and spectral properties of these graphs. We explore the low-energy behavior of these models in Section~\ref{sec:phases}, characterizing their low-energy phases and universality, response to quantum fluctuations, and spatial correlations. 

Finally, in Section~\ref{sec:robustness}, we analyze sensitivity of ground-state properties away from ``special'' system sizes. Since experiments will not always realize system  sizes that are  powers of $p$ or Fibonacci numbers exactly, we determine which conclusions survive under changes in $N$. 

Together, these results provide a systematic understanding of how sparse long-range connectivity shapes many-body behavior in both spin and fermionic settings, and allows us to characterize universal features within this family of sparse graphs.

\section{Graphs \& Models}\label{sec:g & m}

In this work, we extend the framework in \cite{gunning_2025} to a broader class of sparse graphs with logarithmically growing degree, focusing on the power-of-$p$ graphs, and the Fibonacci graph. The geometry of these models is most naturally understood by representing spins as nodes of a weighted graph, where interaction strengths are encoded in the edge weights. In this picture, stronger couplings draw nodes closer together, while weaker couplings push them further apart, providing an intuitive picture of the interaction structure. As the tuning parameter $s$ is varied, this representation reveals a continuous deformation of the effective geometry of both graphs as shown in Fig.~\ref{fig:effective geometries}.

\subsection{Our graphs}\label{sec:graphs}

In the \textit{power-of-$p$ graph} $G_p$, an edge is present between two vertices $i,j$ if their separation $d = |i-j|$ belongs to the distance set,
\begin{equation}
    d \in \mathscr{D}_p = \left\{p^l \mid l =0,1,\dots,\log_p\frac{N}{p}-1\right \}.
\end{equation}

The distance set $\mathscr{D}_p$ enforces sparsity in the graph connectivity, constraining edges to exist between nodes separated by powers of $p$. Unless otherwise stated, we will consider system sizes of the form $N = p^l$, which ensures consistent scaling. This construction has a natural connection to $p$-adic number theory, which will be leveraged in the analysis below.

In contrast, we also consider a graph defined by Fibonacci distances as it breaks the discrete scale symmetry present in power-$p$ models (for which $d \in \mathscr{D}_p$ implies $pd \in \mathscr{D}_p$), which can introduce structural constraints on the many-body Hilbert space. The \textit{Fibonacci graph} $G_{\rm fib}$, is defined by placing an edge between vertices $i,j$ if their separation belongs to,
\begin{equation}
    d \in \mathscr{D_{\rm fib}} = \left\{F_n:F_n \leq \frac{N}{2}\right \},
\end{equation}
where $\{F_n\}$ denotes the Fibonacci sequence, defined by $F_1=1$, $F_2=2$, and $F_{n+2}=F_{n+1}+F_n$. Since the Fibonacci numbers scale exponentially, $\mathscr{D_{\rm fib}}$ again generates the desired sparsity between nodes separated by Fibonacci numbers, but without the self-similarity symmetry from the power-of-$p$ graphs. Unless otherwise stated, we will consider system sizes $N \in\{ F_n\}$, which ensures consistent scaling.

\begin{figure*}[t]
    \centering
    \includegraphics[width=0.99\linewidth]{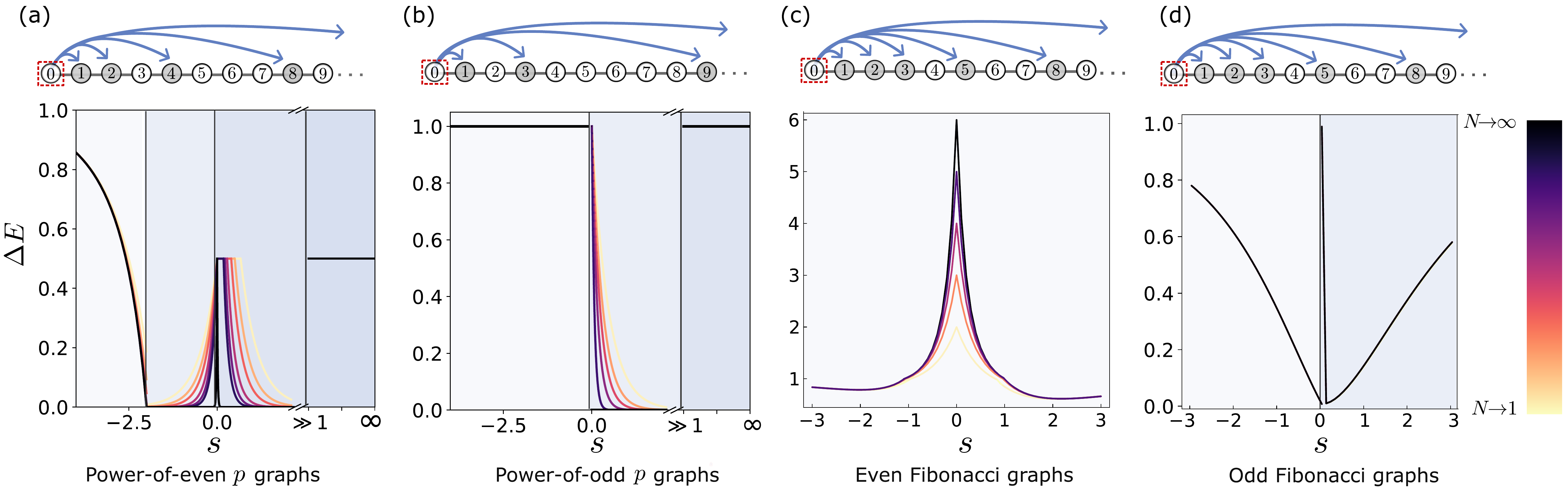}
    \caption{\textit{Energy gap analysis characterising phases of the classical Ising model on power-of-$p$ and Fibonacci graphs.}
(a) Power-of-two phase diagram, representative of all even $p$. Four distinct phases appear in the thermodynamic limit. As $p$ increases, the first critical point (located at $s=-2$ for $p=2$) shifts towards $s=-1$.
(b) Power-of-three phase diagram, representative of all odd $p$. Three phases occur in the thermodynamic limit. The antiferromagnetic (AFM) configuration remains a ground state for all $s$; for $s>0$, an extensive manifold of competing states becomes degenerate with the AFM.
(c) Fibonacci graph for even Fibonacci system sizes $N_{\rm fib}=34,144,610,\dots$. No phase transition is observed: the AFM remains the ground-state for all values of $s$.
(d) Fibonacci graph for odd Fibonacci system sizes $N_{\rm fib}=13,21,55,\dots$. Two phases occur. For $s<0$ the AFM is the ground state, while for $s>0$ a distinct configuration that better satisfies the dominant furthest-neighbour interactions becomes energetically favoured.}
    \label{fig:phases}
\end{figure*}

\subsection{Our models}\label{sec:our models}

These graphs are now used as interaction graphs in our many-body Hamiltonians. We explore both the transverse field Ising model and a fermionic hopping model on each of our sparse graphs. In both models, interactions between spins or fermions are defined as a distance-dependent strength $J_d=Jd^s$ where the exponent $s$ acts as a control knob for the effective interaction range. For all graphs, the limit $s\rightarrow -\infty$, reduces interactions to local, nearest-neighbor interactions. As $s\to 0^-$, the relative dominance of nearest-neighbor bonds weaken. At $s=0$, all bonds (edges) present contribute equally. For $s>0$, the dominant interactions shift toward the largest distances allowed by $\mathscr{D}_G$.  A renormalization factor $(2/N)^s$ is introduced in the $J_d$ term here to constrain interaction strength to a maximum value $J$. In power-of-$p$ graphs, this regime admits a natural interpretation in terms of $p$-adic geometry, where locality is more naturally described on a tree-like structure. The exponent $s$ therefore acts as a control parameter governing the low-energy phases of the system.
The transverse field Ising Hamiltonian on a graph $G$ takes the form,
\begin{equation}\label{eq:H_IM}
    H_{TFIM}^G =  \sum_{i=1}^N\sum_{d \in \mathscr{D}_G}J_{d} \; S_i^zS_{i+d}^z\; + B \sum_i^N S_i^x ,
\end{equation}
where $B$ is the transverse field. We consider a spin-$1/2$ model, where $S_{i}^{\alpha}=\sigma_{i}^{\alpha}/2$, with $\hbar = 1$ and $\alpha \in \{x,z\}$ with periodic boundary conditions imposed, $S_{i+d}^\alpha \equiv S_{(i+d \mod N)}^\alpha$.

The second model describes spinless fermions with long-range hopping. The Hamiltonian is given by, 
\begin{equation}\label{eq:H_F}
    H_{FH}^G =  \sum_{i}^N\sum_{d \in \mathscr{D}_G}J_{d} \; c_i^{\dagger}c_{i+d}\; + \Delta \sum_i^N c_i^{\dagger}c_{i+1}^{\dagger} +\text{h.c.},
\end{equation}
where $c_i$ is a fermionic annihilation operator, and $\Delta$ denotes the pairing strength.\\

\section{Phases, fluctuations and Correlations}\label{sec:phases}

\subsection{Transverse field Ising model on power-of-p graphs}\label{sec:TFIM-pwrp}

\subsubsection{Classical limit $B=0$}\label{sec:classical ising}
We start by exploring the transverse-field Ising model on the power-of-$p$ graph in the limit $B=0$ and antiferromagnetic couplings $J>0$ in Eq.\eqref{eq:H_IM}. As in \cite{gunning_2025}, the classical phase diagrams are determined numerically using exact diagonalization for small system sizes, Monte Carlo methods for larger systems, and analytical proofs that access the thermodynamic limit directly. Detailed numerical and analytical calculations are presented in Appendix~\ref{app:classical}. Together these methods determine the classical phase diagrams and provide a consistent picture across all system sizes. The geometric arguments included below additionally offer an intuitive understanding of the resulting phases and critical behaviour.

\begin{figure}[t]
    \centering
    \includegraphics[width=0.99\linewidth]{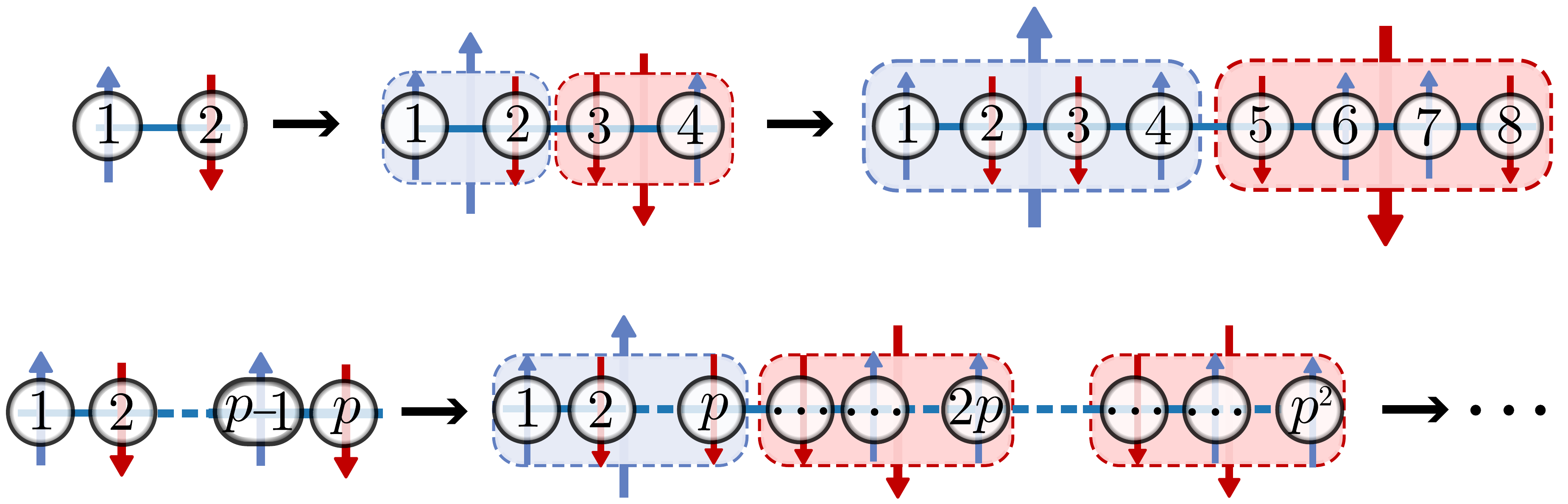}
    \caption{\textit{Recursive construction of ground-state configurations for power-of-even $p$ graphs.}
For $p=2$ (top), the system size is doubled by appending a globally flipped copy of the previous configuration, yielding an antiferromagnetic ordering of two effective blocks (shown in green/red). Iterating this step generates larger system sizes.
For general even $p$ (bottom), the construction proceeds analogously by concatenating $p$ copies of the previous configuration with alternating sign, producing an antiferromagnetic arrangement of $p$ effective blocks.}
    \label{fig:even_p}
\end{figure}

\noindent\textit{$p=2$}: The low-energy phase diagram for the power-of-two graph is characterized in \cite{gunning_2025} and can be seen here in Fig.~\ref{fig:phases}(a). The ground-state is the antiferromagnet (AFM) for sufficiently negative $s$, and the lowest excitation consists of a pair of domain walls. At a critical value $s_c=-2$, the AFM ceases to be the ground-state and the system enters a gapless regime. Upon further increasing $s$, a gap reopens and the ground-state is replaced by a configuration that can be constructed recursively as the system size is doubled, as shown in Fig.~\ref{fig:even_p} (top). Starting from the ground-state at a given system size, all spins in this configuration are treated as a single effective block. The spins in this block are then globally flipped and appended to the original configuration, doubling the system. This construction yields an antiferromagnetic arrangement of two effective blocks and can be iterated to generate the ground-state at larger system sizes by forming a longer effective AFM.\\

\noindent\textit{$p$ even:} The behavior observed for the power-of-two graph extends directly to all even values of $p$. For sufficiently negative $s$, the ground state is the AFM, which loses stability at a first critical point $s_c(p)$. As in the $p=2$ case, this is followed by an intermediate gapless regime, and upon further increasing $s$ a gap reopens with a ground-state that can be constructed recursively by grouping spins into blocks of size $p$ and alternating the overall sign of successive blocks. This construction is identical to the $p=2$ case, with the sole modification that the system size grows by a factor of $p$ at each step rather than by a factor of two. This recursive configuration is shown explicitly in Fig.~\ref{fig:even_p} (bottom) and further details are provided in Appendix~\ref{app:even_p_gaps}.

The qualitative structure of the phase diagram is therefore the same for all even $p$. The primary effect of increasing $p$ is to shift the first critical point $s_c(p)$ toward $s=-1$ (shown in Appendix), reflecting the fact that the next competing interaction after nearest neighbors ($d=1$) occurs at an increasing distance ($d=p$) and nearest-neighbors dominate over a wider range of $s$. In Appendix~\ref{app:even_p_gaps}, this is shown for the cases $p=4,\;6$. \\

\noindent\textit{$p=3$:} The power-of-three graph introduces an important new source of frustration compared to the $p=2$ case. Each interaction distance $d=3^i$ decomposes the ring into $\gcd(N,d)=d$ disjoint cycles, each of length $N/d=3^{L-i}$. Since these cycles all have odd length, an antiferromagnetic configuration cannot satisfy all bonds within any cycle, leading to frustration.

\begin{figure}[b]
    \centering
    \includegraphics[width=0.99\linewidth]{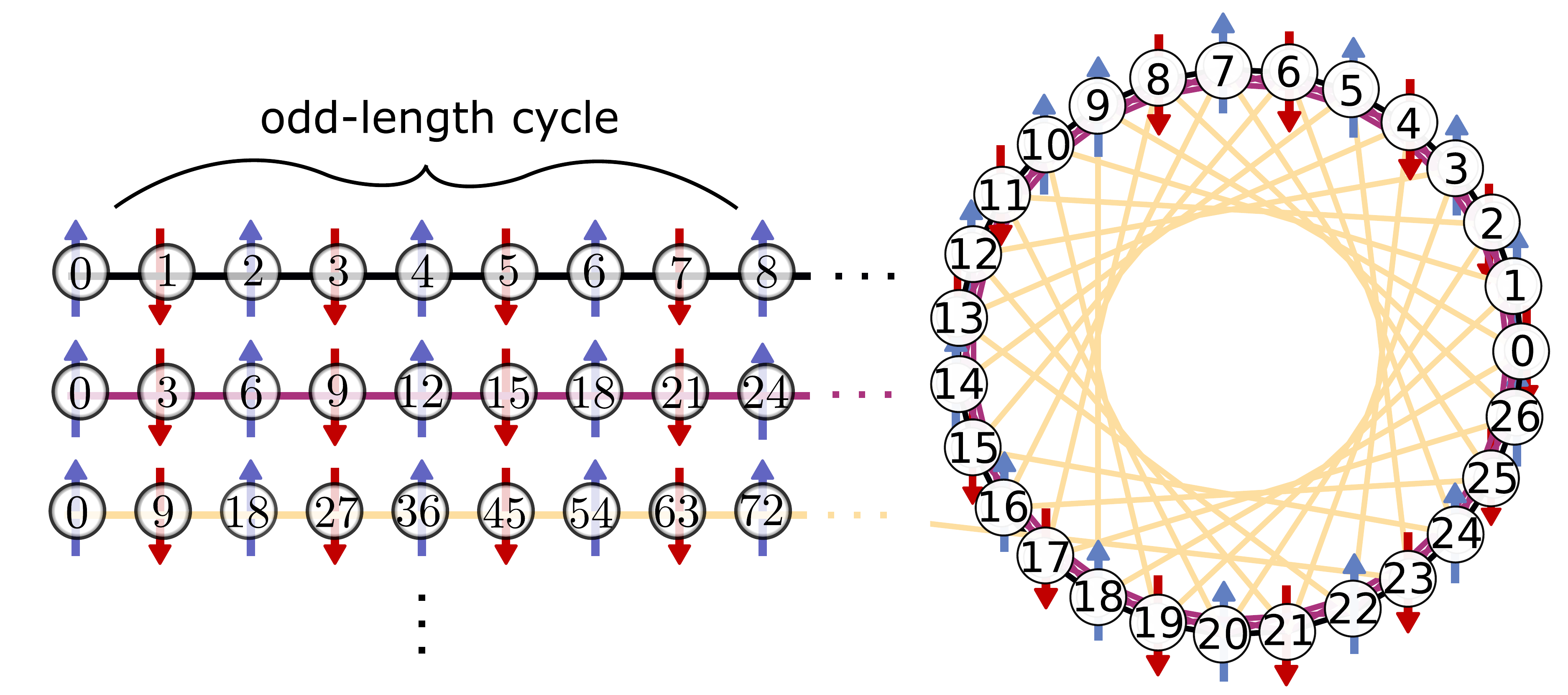}
    \caption{\textit{Antiferromagnetic state on the $p=3$ graph revealing minimal frustration.}
    (Right) The power-of-three graph with each interaction distance $d=3^i$ generating a closed cycle on the ring. Cycles corresponding to $d=1$ (black), $d=3$ (purple), and $d=9$ (peach) are shown. Spins on odd sites are up (blue) and spins on even sites are down (red), corresponding to the antiferromagnetic configuration.
    (Left) Each colored cycle is unraveled into a one-dimensional chain. Since all cycles have odd length, an antiferromagnetic ordering necessarily frustrates one bond per cycle, visible at the chain ends. This is the minimal number of frustrated bonds possible on an odd-length cycle.}
    \label{fig:odd_p}
\end{figure}

Despite this frustration, the one-domain-wall AFM configurations remain the global ground-states for all values of $s$. These configurations frustrate \emph{exactly one bond} on each cycle associated with $d=3^i$, which is the minimal number of frustrated bonds possible on an odd-length cycle. This minimal frustration is shown explicitly in Fig.~\ref{fig:odd_p}. Any deviation from this AFM manifold necessarily introduces additional frustrated bonds on at least one of the cycles generated by the distances $3^i$. Since all such cycles are present simultaneously in the power-of-three graph, the AFM configurations minimize the total number of unsatisfied bonds across all length scales. In the positive-$s$ regime, however, the longest-distance shell increasingly dominates the energy. Configurations that optimize this dominant shell but differ on shorter-distance shells can therefore become asymptotically degenerate with the AFM manifold as the short-range length scales become energetically irrelevant.

The resulting phase diagram, shown in Fig.~\ref{fig:phases}(b), is therefore surprisingly simple: the antiferromagnetic state remains the ground state throughout the entire parameter range. For $s>0$, excited states that maximize satisfaction of the longest-range bonds while breaking shorter-range couplings approach the AFM in energy, leading to a collapse of the excitation gap and a gapless regime. In the limit $s\to\infty$, these states become degenerate, with a finite gap separating them from configurations that frustrate an additional longest-range bond.\\
\noindent\textit{$p$ odd:}
The arguments for $p=3$ generalize to all odd $p$. Since all distances $d=p^i$ are odd, each generates an odd-length cycle on the ring. The AFM state frustrates exactly one bond per cycle, which is minimal. Consequently, all odd-$p$ graphs share the same qualitative phase diagram, as shown in Appendix~\ref{app:odd_p_gaps}.

\subsubsection{Quantum model $B\neq0$}\label{sec:quantum pwrp}
We now extend to $B \neq 0$ in Eq.~\eqref{eq:H_IM} to determine how a transverse field will modify the phases established above.

\begin{figure}[t]
    \centering
    \includegraphics[width=0.99\linewidth]{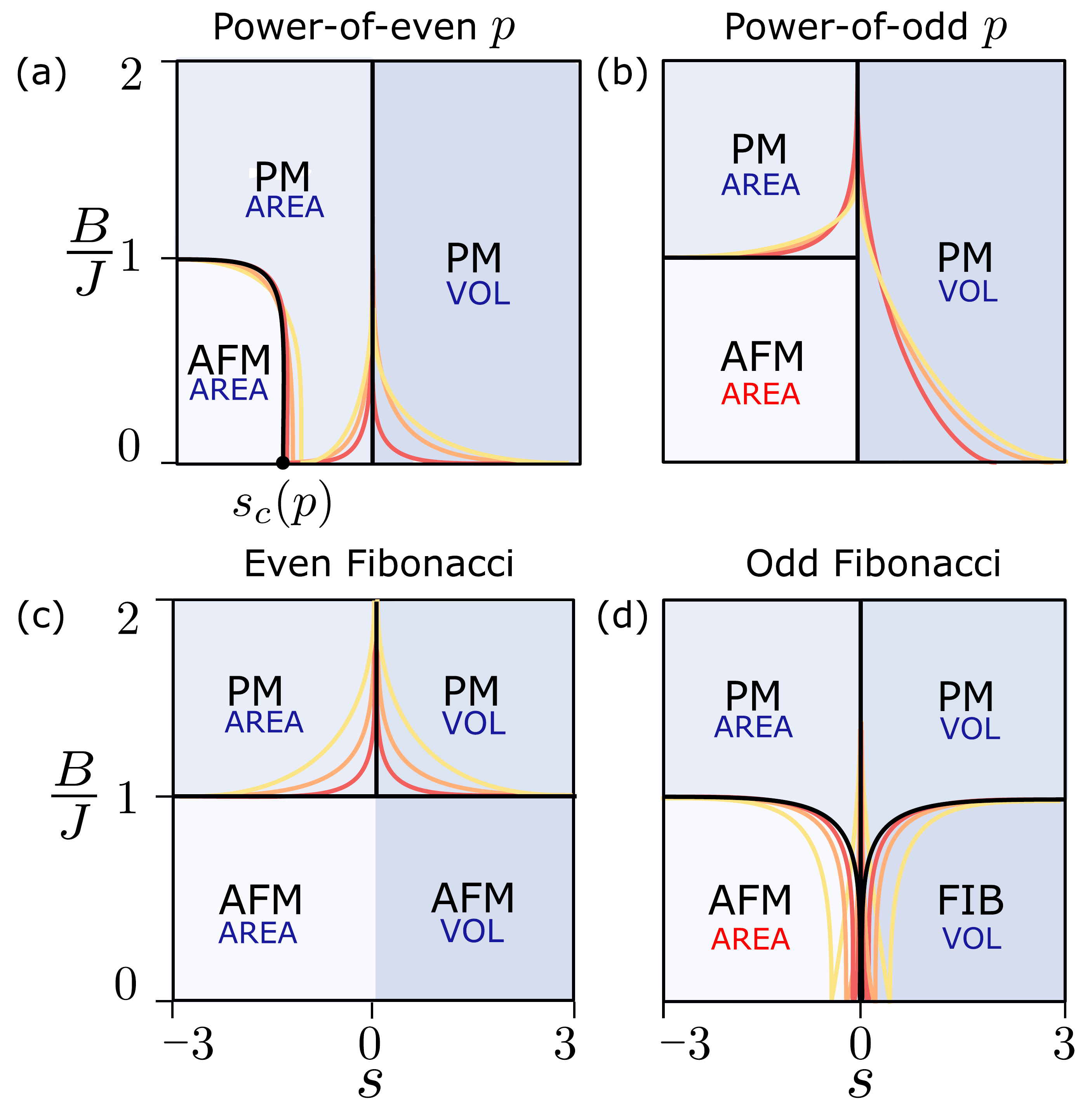}
    \caption{\textit{Schematics of quantum phase diagrams for the transverse-field Ising model on logarithmically sparse graphs.} Panels (a) and (b) show representative power-of-$p$ graphs: $p=2$ for the even-$p$ class with $N=8,16,32$ and $p=3$ for the odd-$p$ class with $N=9,27,81$. Panels (c) and (d) show Fibonacci graphs for even and odd Fibonacci system sizes, respectively, using the accessible sequences $N=8,34,144$ and $N=13,21,55$. Colored curves denote finite-size estimates of phase boundaries where numerical convergence is obtained, while the bold black curves summarize the thermodynamic extrapolations, combining finite-size scaling, SKQD results, and analytically controlled limiting regimes. Shaded regions are labeled by phase (antiferromagnet, AFM; Fibonacci antiferromagnet, FIB; paramagnet, PM) and annotated by their entanglement scaling (area-law vs volume-law) with respect to partitions in the original site ordering. The classical critical line $s_c(p)$ is plotted for power-of-even $p$ graphs; it approaches $s=-1$ as $p$ increases.}
    \label{fig:quantumphases}
\end{figure}

\noindent\textit{$p$ even:} In \cite{gunning_2025}, the quantum phase diagram of the power-of-two graph was characterized in detail and is reproduced in Fig.~\ref{fig:quantumphases}(a). The classical structure identified above strongly constrains the quantum phases for even $p$. Consequently, the quantum phase diagrams for all even $p$ are expected to agree qualitatively with Fig.~\ref{fig:quantumphases}(a). In Appendix~\ref{app:even_p_quantum}, we explicitly verify this agreement for $p=4$, with $N=16$ across all regimes and for $N=64$ in selected limits. Extending this analysis to larger $N$ or $p$ would require system sizes beyond the reach of our tensor-network convergence.

\noindent\textit{$p$ odd:} Direct tensor-network simulations with periodic boundary conditions and long-range couplings become computationally prohibitive for large systems. For the representative case $p=3$, proper finite-size scaling requires system sizes $N=9,27,81$, but reliable convergence is not achieved for $N=81$ within our current framework. Nevertheless, several controlled limits of the model remain fully accessible and impose strong constraints on the global structure of the phase diagram. By combining these limits with finite-size density matrix renormalisation method (DMRG) with matrix product states (MPS) \cite{WhiteFeiguin2004, DaleySchollwoeck2004, Schollwoeck2011, FishmanStoudenmire2022, ITensor2022} results where available, we construct a consistent picture of the quantum phase structure. We now layout our methods to scale to larger system sizes in particular regimes of interest.

In the limit $s \to -\infty$, the Hamiltonian reduces to the nearest-neighbour transverse-field Ising model, whose quantum critical structure is exactly known. In the opposite limit $s \to +\infty$, the model reduces to a furthest-neighbour model, which is again analytically tractable. For $B \gg 1$, the transverse field dominates and the system enters a trivial paramagnetic phase.

A remaining non-trivial limit corresponds to $B \ll 1$, where quantum fluctuations weakly perturb the classical ground-state manifold. In this regime, the problem reduces to understanding how the transverse field lifts classical degeneracies and generates effective dynamics within the low-energy manifold. To access this regime, we employ a Krylov- and sampling-based diagonalization strategy, which captures the leading quantum corrections without requiring full tensor-network convergence.

 \textit{Krylov \& Sample based quantum diagonalization.} When a small transverse field $B$ is introduced, the quantum ground state remains closely tied to the classical structure. For sufficiently small $B$, the dominant contributions arise from the classical ground states together with a limited number of low-energy spin-flip excitations. This observation allows us to work within a substantially reduced Hilbert space while still accurately capturing the relevant quantum physics.

We exploit this using two complementary approaches. In sample-based quantum diagonalization (SQD), the full Hamiltonian is projected onto a reduced subspace chosen to approximate the ground state \cite{kanno2023,Robledo_Moreno_2025,kaliakin2024accuratequantumcentricsimulationssupramolecular,Barison_2025,liepuoniute2024quantumcentricstudymethylenesinglet,shajan2024quantumcentricsimulationsextendedmolecules}. While selecting an appropriate subspace can be challenging in general, in our case the low-energy classical states have already been identified in Section~\ref{sec:classical ising} and naturally form the core of this reduced basis at each value of $s$. We also employ Krylov quantum diagonalization (KQD), which instead constructs the relevant subspace dynamically by repeatedly applying the Hamiltonian to an initial state \cite{parrish2019quantumfilterdiagonalizationquantum,Motta_2019,Stair2020MultireferenceQuantumKrylov,Urbanek2020VirtualQuantumSubspaceExpansion,Cohn2021QuantumFilterDiagonalization,Seki2021QuantumPowerMethod,Baker2024BlockLanczosExcitedStates,Baker2021LanczosRecursionGreenFunctionGroundState,Klymko2022RealTimeEvolutionUltracompactEigenstates,Jamet2022QuantumSubspaceExpansionGreensFunctions,Lee2024SamplingErrorAnalysisQKSD,Kirby2024AnalysisQuantumKrylovErrors,Kirby2023ExactEfficientLanczos,Shen2023RealTimeKrylovTheory,Tkachenko2024QuantumDavidsonExcitedStates,Epperly2022TheoryQuantumSubspaceDiagonalization,Yoshioka_2025}.
A natural choice for this initial state is one of the classical ground states, which greatly simplifies the procedure: the longitudinal $S^zS^z$ Hamiltonian terms act diagonally on classical product states, while the transverse-field term generates the nontrivial basis states through single- and few-spin flips. Combining elements of both methods, we build a reduced subspace consisting of the classical states and their lowest-energy excitations, and diagonalize the Hamiltonian within this space \cite{SKQD_org_IBM}. This allows us to accurately compute both local observables and nonlocal quantities, such as the entanglement entropy, for small $B$.

 In the present case, the number of classical ground-states in the limiting regimes is two (Néel state), and the inclusion of low-energy spin-flip excitations leads to a basis whose dimension scales as $\mathcal{O}(N^2)$. This dramatic reduction from the full $2^N$ Hilbert space allows us to access system sizes up to $N\sim200$, well beyond the range where standard exact diagonalization or DMRG methods are effective for this model.

For the $p=3$ graph, our conclusions are supported by three complementary approaches:
\begin{itemize}
    \item[(i)] DMRG calculations for $N=9$ and $N=27$ across the full $(B,s)$ parameter space,
    \item[(ii)] sample-based and Krylov  (SKQD) in the low-$B$ regime, and
    \item[(iii)] exact results in the asymptotic limits $s \to -\infty$, $s \to +\infty$, and $B \gg J$.
\end{itemize}

Taken together, these are consistent with the quantum phase diagram shown in Fig.~\ref{fig:quantumphases}(b).

The classical $p=3$ model exhibits a single phase transition at $s=0$, which lies along the $B=0$ axis of the quantum phase diagram. We now analyse how quantum fluctuations modify the critical behaviour. For $s<0$, we recover the standard transition between antiferromagnetic and paramagnetic phases at $B=J$ in the nearest-neighbor limit. For sufficiently negative $s$, the transition lies in the two-dimensional Ising universality class. Whether this universality persists as $s \to 0^{-}$ remains unclear, as long-range bonds become increasingly relevant and the system is no longer analytically tractable.

In the nearest-neighbour limits, both phases exhibit area-law entanglement scaling in the thermodynamic limit. For the $p=3$ graph, odd system sizes under periodic boundary conditions necessarily introduce frustration, preventing a clean numerical observation of the asymptotic scaling at accessible system sizes. The area-law regime is therefore indicated in red in Fig.~\ref{fig:quantumphases}(b), and the asymptotic scaling is justified in Appendix~\ref{app:odd_p_quantum}. The $s<0$ region is primarily determined using methods (i) and (iii), with corroborating numerical evidence included in the Appendix.

At $s=0$, methods (i) and (ii) indicate that the singular critical point of the classical phase diagram extends into a quantum critical line at finite $B$.

For $s>0$, methods (i) and (iii) consistently indicate the emergence of a volume-law paramagnetic phase for any $B>0$ in the original site ordering, reflecting the dominance of long-range connectivity in this regime and the gapless behavior of the classical model.

\subsection{Transverse field Ising model on the Fibonacci graph}\label{sec:tfim-fib}

We now consider the transverse-field Ising model on the Fibonacci graph, which we introduce to remove the discrete self-similarity present in the power-of-$p$ graphs and to explore a broader class of sparse interaction geometries. As shown in Fig.~\ref{fig:effective geometries}, tuning the interaction exponent $s$ for power-of-$p$ graphs drives a change from a single ring geometry ($s\ll0$) to $N/p$ decoupled loops of size $p$ ($s\gg0$). Instead, for the Fibonacci graph we find that the geometry remains ring-like in both limits. As a result, there is no obvious source of an extensive ground-state degeneracy for $s>0$, and one expects a somewhat symmetric gapped behaviour on both sides of the phase diagram. 

A distinctive feature of the Fibonacci graph is that it naturally includes both even and odd system sizes (e.g. $N_{\rm fib}=8,13,21,34,\dots$) as the graph is scaled. This is important because antiferromagnetic (AFM) order cannot be perfectly satisfied on an odd-length ring, leading to immediate frustration. We therefore treat even and odd system sizes separately.

\subsubsection{Classical limit $B=0$}

For $s \ll 0$, the exact ground state is the Néel antiferromagnetic configuration $S_j =(-1)^j,$ which identically satisfies the nearest-neighbour condition $S_{j+1} = -S_j$. We must establish whether this state remains energetically optimal once other bonds $d \in \mathscr{D}_{\rm fib}$ are introduced.

If the antiferromagnetic pattern is preserved across bonds of distance $d$, i.e. $ S_{j+d} = -S_j,$ then it is plausible that the Néel state remains optimal even in the presence of such couplings. This question becomes particularly relevant for the furthest-neighbour distance $d_f$, which dominates the Hamiltonian for $s>0$. If the Néel state fails to minimise frustration for bonds of length $d_f$, then one expects a transition away from this configuration when those bonds become dominant.

Under periodic boundary conditions, the spin at distance $d$ from site $j$ is
$ S_{(j+d)\bmod N} = (-1)^{j+d-WN} = S_j\,(-1)^{d-WN},$ where $W \in \{0,1\}$ accounts for wrap-around across the boundary. The condition for the Néel structure to satisfy $S_{(j+d)\bmod N} = -S_j$ is therefore
$(-1)^{d-WN} = -1,$ which requires
$ d - WN \;\text{to be odd}.$ For even Fibonacci system sizes $N_{\rm even}=F_k$, the furthest-neighbour distance is $d_f=F_{k-2}$, which is always odd, since any even Fibonacci number is preceded by two odd ones. Moreover, for even $N$, the wrap-around term $WN$ does not change whether the exponent is even or odd. Consequently, $(-1)^{d_f-WN}=(-1)^{\text{odd}}=-1$, and the Néel AFM satisfies all furthest-neighbour bonds exactly. Since the nearest-neighbour and furthest-neighbour limits favour the same AFM ordering, the limiting geometries provide no mechanism for a change of ground-state order as $s$ is tuned. This expectation is confirmed by our numerical results: as shown in Fig.~\ref{fig:phases}(c), the excitation gap for $N_{\rm even}$ remains finite for all $s$, and no phase transition occurs.

For odd Fibonacci system sizes $N_{\rm odd}$, the situation is qualitatively different. While the Néel state remains the ground state for $s<0$, where nearest-neighbour couplings dominate and only a single bond is frustrated, the furthest-neighbour regime $s>0$ cannot be satisfied in the same way. In this limit the dominant interactions occur at distance $d_f$, and repeated periodic wrap-around along this cycle necessarily flips parity. Consequently, the condition that $d - WN$ is odd
cannot be satisfied uniformly for all bonds in the cycle, and the Néel configuration no longer maximally satisfies the dominant interactions.

This leads to a phase transition at $s=0$ into a new ground state that optimally satisfies the furthest-neighbour couplings. Since the effective geometry in this limit is again ringlike, the resulting ground state is the ``Fibonacci AFM": an antiferromagnetic pattern defined on the ring formed by sites connected at distance $d_f$. In other words, the AFM ordering is preserved not in the nearest-neighbour sense, but along the effective furthest-neighbour cycle.

The resulting phase diagram, shown in Fig.~\ref{fig:phases}(d), exhibits a single transition at $s=0$ and gapped phases on either side.

\subsubsection{Quantum model $B\neq 0$}

We now introduce quantum fluctuations into the Fibonacci model for system sizes $N$ that are Fibonacci numbers. As in the classical analysis, understanding the structure of the ground-state manifold and the effective interaction geometry is essential for interpreting the quantum behaviour. To this end, we employ the three techniques introduced in Sec.~\ref{sec:quantum pwrp}: (i) DMRG, (ii) SKQD, and (iii) analytical methods in the limits.

For the Fibonacci model, method (ii) proves particularly powerful. The key geometric property is that both the nearest-neighbour distance $d=1$ and the furthest allowed Fibonacci distance $d_f$ generate a single connected cycle on the lattice. Consequently, in both asymptotic limits of the model the interaction graph reduces to a single ring, leading to a unique classical ground state (up to a global spin flip). There is therefore no extensive degeneracy in either limit.

The situation changes only near $s=0$, where interactions at multiple length scales compete and the manifold of low-energy states expands. As the efficiency of SKQD depends directly on the size of the relevant low-energy subspace, this method performs particularly well in the two regimes with unique and gapped ground states, allowing us to access system sizes far beyond those currently achievable with MPS methods. This behaviour contrasts with the power-of-$p$ graphs discussed previously, where gapless regions appear in both limits. In those cases, the low-energy manifold remains extensive, limiting the efficiency gains of SKQD and requiring it to be used primarily as a complementary benchmark to DMRG.
The phase diagrams obtained from these three complementary approaches are mutually consistent and are summarized in Fig.~\ref{fig:quantumphases} (c) and (d).

For odd Fibonacci system sizes, the introduction of quantum fluctuations gives rise to four phases, as shown in Fig.~\ref{fig:quantumphases}(d). In the classical limit, two phases were identified: the Néel antiferromagnet for $s<0$ and the Fibonacci antiferromagnet for $s>0$. Importantly, the effective interaction geometry in both limits is identical — 1D ringlike — differing only in which bonds define the antiferromagnetic ordering. As a consequence, the quantum phase diagram exhibits a symmetry across the $s=0$ line. For $s \ll 0$, the system undergoes a transition at $B=J$ from the Néel antiferromagnet to a paramagnetic phase. Similarly, for $s \gg 0$, there is a transition at $B=J$ from the Fibonacci antiferromagnet to a paramagnet. In both limits, the effective interaction geometry reduces to a one-dimensional ring, and the transition lies in the 1D quantum Ising universality class (equivalently, the 2D classical Ising class). This is a particularly striking result: despite the $s>0$ limit corresponding to a fully long-range model, the effective interaction geometry reduces to a one-dimensional ring (shown in Fig.~\ref{fig:effective geometries}). This reduction is not evident from the interaction profile alone and highlights the importance of the geometry-based perspective in understanding the emergent many-body physics and transitions. Approaching $s=0$ from either direction, the classical spectrum becomes gapless as competing interaction scales render the low-energy manifold increasingly dense. The energy cost of accessing excitations is therefore reduced, and even small transverse fluctuations can efficiently mix states. As a result, the critical transverse field required to destabilize the ordered phase is suppressed, with the transition line bending toward $B \to 0$ as $s \to 0$. Detailed numerical results are presented in Appendix~\ref{app:odd_fib_quantum}.

For even Fibonacci system sizes, the quantum phase diagram is an extension of a trivial classical phase diagram with one phase. As for $B=0$ there was no phase transition for any $s$ and the AFM remains the ground-state. Therefore the main transition in both limits is the AFM to paramagnet occuring at $B=J$ and as $s\to0$ a larger transverse field ($B>J$) is required to destabilise the AFM as all bonds present in the fibonacci graph contribute equally to fight against the quantum fluctuations. The same symmetry on either side of $s=0$ is found again here due to the underlying geometry. Numerical results are laid out in Appendix~\ref{app:even_fib_quantum} and a schematic of the phase diagram is shown in Fig.~\ref{fig:quantumphases}(c).

\subsection{Fermionic hopping model on the power-of-p graphs}\label{sec:Fhm-pwrp}

\begin{figure}[t]
    \centering
    \includegraphics[width=\linewidth]{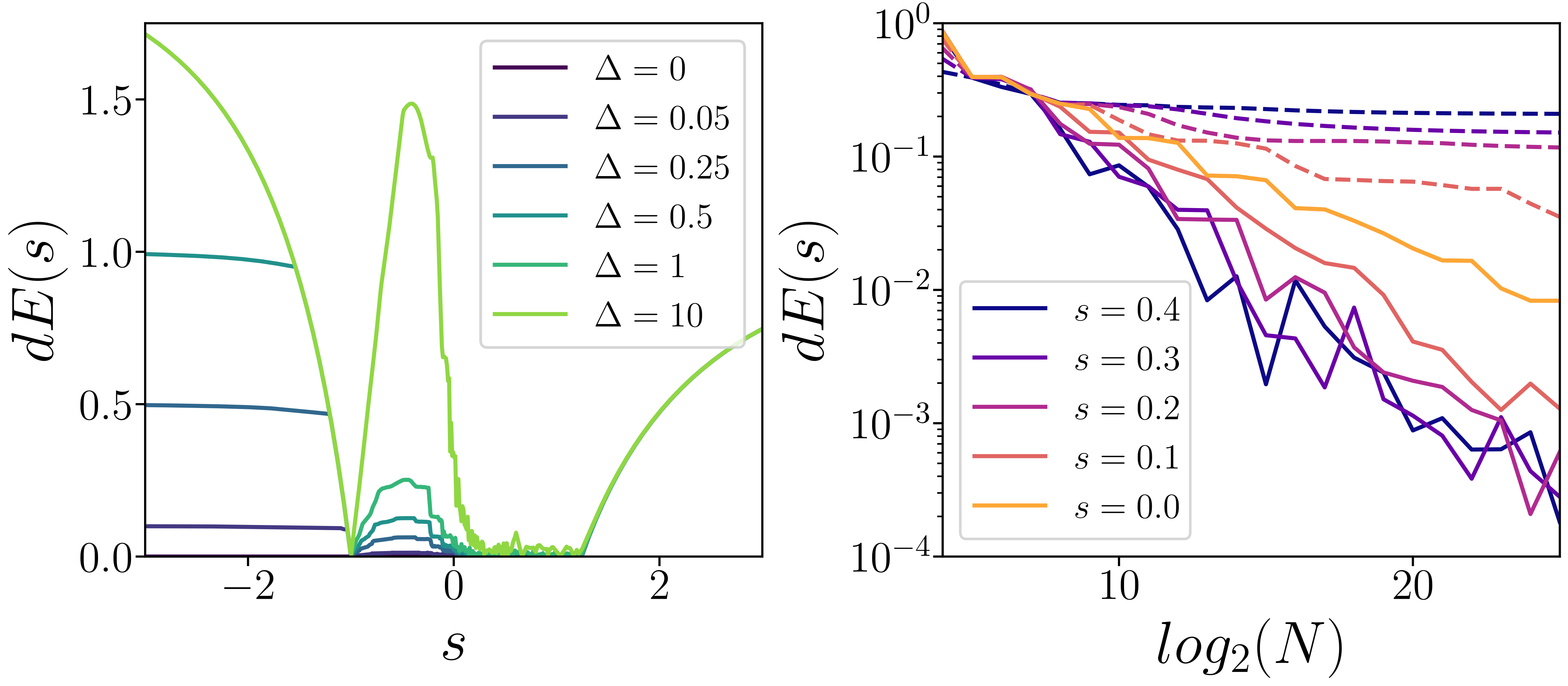}
    \caption[PWR2 spectral gap and spectrum]{\textit{Spectral gap and coupling-function $\mathscr{J}_k$ of the PWR2 model for different parameters.} The gap closes analytically at $s=-1$ as shown in Eq.~\eqref{eq: pwr2_gapcloser}. This is corroborated numerically in (a), where the spectrum is well converged with system size and remains gapped except at the critical point for $s<0$. For $s>0$, the analytic argument no longer determines the true gap minimum; numerical results instead show a second gapless regime extending from $s=0$ to $s^+\approx 1.2$. (b) Gap scaling in this positive-$s$ gapless regime. The dashed lines indicate the corresponding values of $-s$. The gap decreases algebraically with system size throughout $0\le s<s^+$, confirming that this region is gapless in the thermodynamic limit.}
    \label{fig:pwr2_spectrum}
\end{figure}

The fermionic hopping model Eq.~\eqref{eq:H_F} shares the same basic structure as the TFIM: one term carries the long-range graph geometry, while the second remains local. Here, the hopping amplitudes $J_d$ are defined on the graph $\mathscr{D}_G$, whereas the pairing strength $\Delta$ acts only between nearest neighbours on the original ring \cite{Buyskikh_2016}. Because the Hamiltonian is quadratic in the fermionic operators, it is exactly solvable through a Fourier transformation followed by a Bogoliubov transformation on independent momentum-space sectors. This brings us from Eq.~\eqref{eq:H_F} to,
\begin{equation}
    \begin{aligned}
        \mathscr{H} &= \frac{1}{2}\sum_{k}
        \begin{pmatrix}
            c_k^{\dagger} & c_{N - k}
        \end{pmatrix}
        \begin{pmatrix}
             \mathscr{J}_k & \overline{\mathscr{D}_k} \\
            \mathscr{D}_k & -\mathscr{J}_k
        \end{pmatrix}
        \begin{pmatrix}
            c_k \\
            c_{N - k}^{\dagger}
        \end{pmatrix}
    \end{aligned}.
    \label{eq:fermion_h_fourier}
\end{equation}
Here $\mathscr{J}_k$ is given by,
\begin{equation}
    \mathscr{J}_k \left( s \right) = \left\{2\sum_{d \in \mathscr{D}_p} J_d \left( s \right) \cos\left( \frac{2\pi k}{N} d \right)\right\} - S_k.
    \label{eq:j_k}
\end{equation}
The sum goes over all distances $d$, for which the coupling-strength is given by $J_d(s)$. For the power-of-$p$ graphs, the antipodal correction $S_k$ contributes only for $p=2$, where $S_k = J_{N/2}\cos(\pi k).$ This accounts for the fact that the furthest distance in the ring-geometry appears without partner, $N/2 \in \mathbb{N}$. For all $p>2$, $S_k=0$.
The fourier-transformation of the nearest-neighbour pairing is,
\begin{equation}
    \mathscr{D}_k = 2 i \Delta \sin \left( \frac{2\pi}{N}k \right),
    \label{eq:pairing_k}
\end{equation}
and the fourier-transformation of the fermionic annihilation operator is,
\begin{equation}
    c_k = \frac{1}{\sqrt{N}}\sum_{r = 0}^{N-1} e^{-\frac{2\pi i}{N} k\cdot r} c_r.
\end{equation}
Diagonalizing the $2\times2$ blocks in Eq.~\eqref{eq:fermion_h_fourier} yields the quasiparticle energies,
\begin{equation}
    \epsilon_k = \sqrt{\lvert \mathscr{J}_k \rvert^2 + \lvert \mathscr{D}_k \rvert^2}.
    \label{eq:fermion_spectrum}
\end{equation}

\subsubsection{Energy-gaps}
The system remains gapped unless a momentum node $k$ exists for which $\epsilon_k = 0$. From Eq.~\eqref{eq:fermion_spectrum}, the gap can close only if both $\mathscr{J}_k=0$ and $\mathscr{D}_k=0$. For $\Delta \neq 0$, the pairing term $\mathscr{D}_k = 2 i \Delta \sin(2\pi k/N)$ vanishes only at $k=0$ and, for even $N$, at $k=N/2$. At $k=0$, however, $\mathscr{J}_0 = 2\sum_d J_d > 0$, so the spectrum remains gapped. The only remaining candidate for gap closing is therefore $k=N/2$, which we analyse using Eq.~\eqref{eq:j_k},
\begin{equation}
    \mathscr{J}_{N/2} = 2 \left\{\sum_{d \in \mathscr{D}_p} J_d \cos{\pi d}\right\} - S_{N/2}.
    \label{eq:j_nhalf}
\end{equation}

\noindent\textit{$p$ even:}
For powers of $p$, the cosine in Eq.~\eqref{eq:j_nhalf} simplifies to $\cos{\pi p^r}$ with $r$ a natural number below $Q=\log_p{N/2}$. For $p$ even, the cosine evaluates to $+1$ except for $r=0$. The calculation is then simple, as shown for $p=2$:
\begin{equation}
    \begin{aligned}
        \mathscr{J}_{N/2} &= 2 \left\{\sum_{q=1}^{Q} J_{2^q}\right\} - 2J_1 - S_{N/2}, \\
        &= \frac{1}{1 - 2^{s}}\left[ 2^{s + 2} - 2 - 2^{s \left( Q + 1 \right)} - 2^{s Q} \right].
    \end{aligned}
    \label{eq: pwr2_gapcloser}
\end{equation}
Taking the thermodynamic limit in $N$, leads to a gap closing at $s = -1$. This agrees with numeric calculations of the energy gap in Fig.~\ref{fig:pwr2_spectrum}(a). One may also see this from considering a truncated version of the model in which only interaction distances up to $p^Q$ are retained, even when $p^Q < N/2$. In this case, $Q$ becomes an independent parameter controlling the number of long-range scales included in the Hamiltonian. For fixed $Q$ the critical exponent $s_{c}$ at which the gap closes is determined by the relation
\begin{equation}
    2^{s_{c} \left( Q + 1 \right)} = 2^{s_{c} + 1} - 1.
\end{equation}
For $s\geq0$, the dominant long-range couplings cause the overall energy scale to diverge with system size, so we instead employ normalization as before. Evaluating the normalized quantity shows that the special momentum mode at $k=N/2$ does not close in this regime.

However, the physical spectral gap is determined by the minimum quasiparticle energy over all momenta, not just the analytically tractable mode at $k=N/2$. While the latter remains finite at $s=0$, the location of the true minimum becomes difficult to determine analytically due to the oscillatory structure of $\mathscr{J}_k$. We therefore compute the full dispersion numerically. The resulting gap is shown in Fig.~\ref{fig:pwr2_spectrum}(a). The scaling in Fig.~\ref{fig:pwr2_spectrum}(b) indicates a gapless thermodynamic regime existing for $s \geq 0$, before the system becomes gapped again beyond a critical value $s^+\approx1.2$ irrespective of the value for $\Delta$. By contrast, the dashed curves for $s<0$ show no corresponding scaling toward zero with increasing system size.

\noindent\textit{$p$ odd:}
For odd $p$, the cosine in Eq.~\eqref{eq:j_nhalf} evaluates to $-1$ for all powers $p^r$, so the analytically tractable mode at $k=N/2$ never closes. For $s<0$, this indicates that the system remains gapped for all values of $s$, in agreement with the numerical results shown in Appendix~\ref{app-ferm-pwr3}. 

For $s\geq0$, the same analytic conclusion holds for the special $k=N/2$ mode. However, as in the even-$p$ case, the physical spectral gap is determined by the minimum quasiparticle energy over all momenta. Numerical evaluation of the full dispersion shows that the gap nevertheless scales toward zero in the thermodynamic limit throughout an extended positive-$s$ regime, before reopening beyond a critical value $s_c$.

\section{Sensitivity away from special system sizes}\label{sec:robustness}

\begin{figure}[t]
    \centering
    \includegraphics[width=1.0\linewidth]{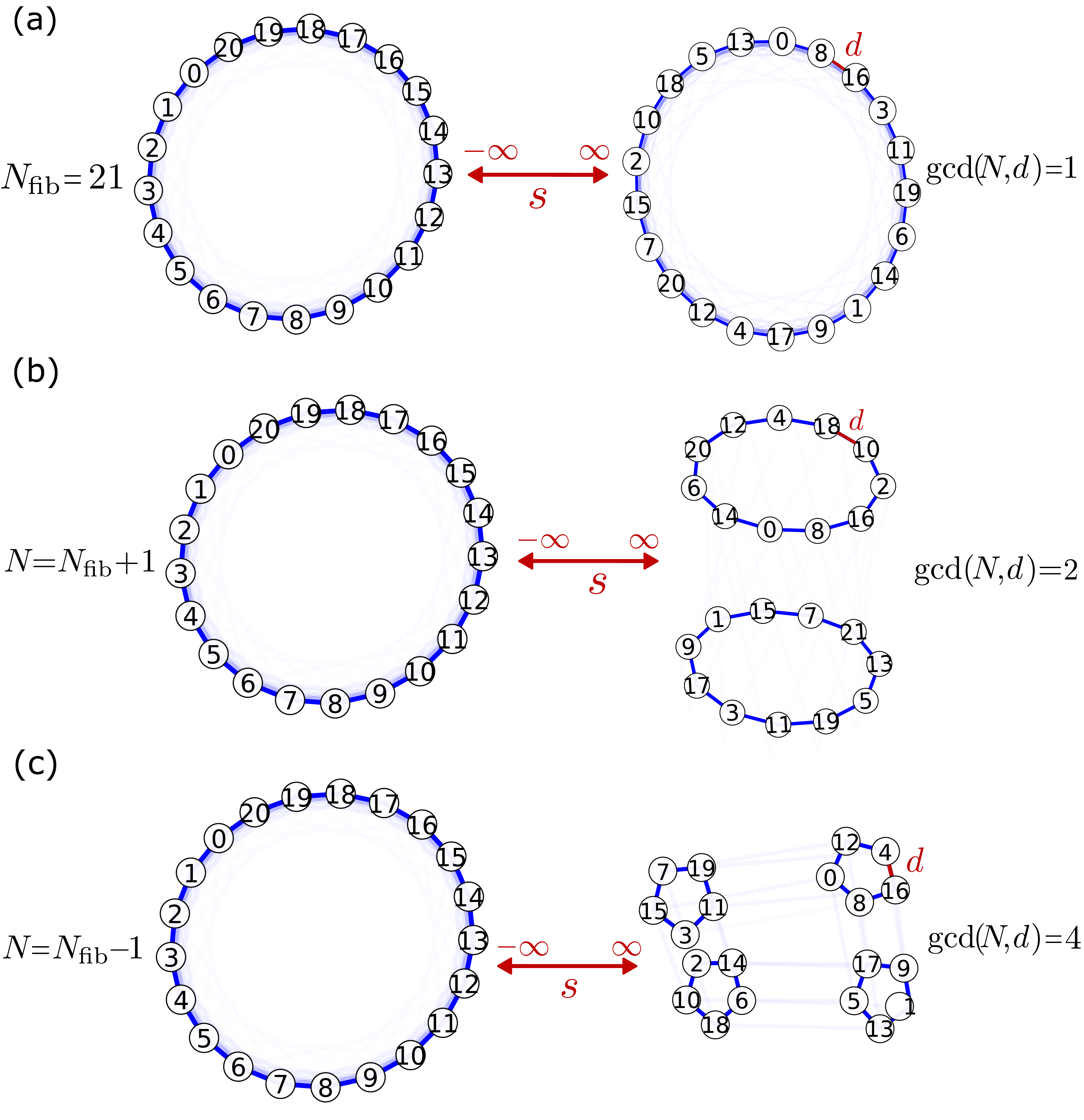}
    \caption{\textit{Effective interaction geometry of the Fibonacci graph on various system sizes.} Weighted interaction graphs where nodes represent spins and edge weights encode the relative interaction strength between spins separated by allowed Fibonacci distances. In each panel, the left image corresponds to the limit $s\to -\infty$, where nearest-neighbour interactions dominate and the geometry is a single ring, while the right image corresponds to the limit $s\to +\infty$, where the dominant interaction connects spins at the furthest allowed distance $d_f$. (a) Fibonacci system size $N_{\rm fib}=21$, for which the dominant long-range interaction connects sites separated by distance $d=8$. Since $\gcd(21,8)=1$, the long-range interactions form a single loop. (b) System size $N=N_{\rm fib}+1=22$, where the same long-range interaction produces two disconnected loops, consistent with $\gcd(22,8)=2$. (c) System size $N=N_{\rm fib}-1=20$, where the long-range interaction produces four disconnected loops, corresponding to $\gcd(20,8)=4$.}
    \label{fig:loops}
\end{figure}

\subsection{Transverse field Ising model}

Throughout this work, we have focused on system sizes belonging to the natural scaling sequences of the graphs under consideration, namely $N=p^l$ for power-of-$p$ graphs and $N\in\{F_n\}$ for Fibonacci graphs. We now relax this restriction and ask how the phase structure changes when the system size is varied away from these special values. This question is particularly relevant for experimental realizations of these models, where precise control over the total number of spins may not always be possible. We find that the qualitative structure of the ground-state phase diagram is entirely governed by the greatest common divisor $\gcd(N,d)$ of the system size $N$ and the interaction distance of interest $d$. In particular, when considering all spins separated by this distance $d$, they partition into $\gcd(N,d)$ disconnected loops, which defines the effective geometry in regimes dominated by that interaction distance.

This observation provides a simple geometric interpretation of several cases discussed earlier. For nearest-neighbour interactions $(d=1)$, one has $\gcd(N,1)=1$, corresponding to a single ring. For power-of-$p$ graphs at the furthest-neighbour scale, with $N=p^i$ and $d=p^{i-1}$, one finds $\gcd(N,d)=p^{i-1}=N/p$, explaining the appearance of $N/p$ disconnected loops in the $s\gg0$ limit, as shown in Fig.~\ref{fig:effective geometries}. This also highlights a key distinction of the Fibonacci graph: for Fibonacci system sizes $N=F_n$, the furthest-neighbour distance $d_f=F_{n-2}$ satisfies $\gcd(N,d_f)=1$. Consequently, the dominant interaction graph in the $s\gg0$ limit remains a single ring, in contrast to the power-of-$p$ graphs where the system fragments into multiple disconnected loops.

Our results can be summarized as follows. For $s<0$, nearest-neighbour interactions dominate and the effective geometry is always a single loop, since $\gcd(N,1)=1$ regardless of system size. Adding or removing a finite number of spins therefore introduces only local changes, and the phase structure for $s<0$ is largely insensitive to variations in $N$. Although the gcd associated with other short-range distances $d=2,3,\dots$ may vary with $N$, it is bounded by $d$ itself, so the resulting loop structure remains minimal and does not qualitatively affect the phase behaviour in this regime.

For $s>0$, the situation is markedly different. In this regime the dominant interactions are furthest-neighbour bonds, and even a change of a single site can qualitatively alter the effective geometry. The structure of the $s\gg0$ regime is fully determined by $\gcd(N,d_f)$, where $d_f$ denotes the furthest-neighbour distance. As shown in Fig.~\ref{fig:loops}(a), If $\gcd(N,d_f)=1$, the effective geometry consists of a single loop and the ground state is unique and gapped for $s>0$. As shown in Fig.~\ref{fig:loops}(b \& c), if $\gcd(N,d_f)>1$, the system decomposes into multiple disconnected loops, leading to an extensive ground-state degeneracy and a gapless manifold. In the power-of-two case, this degeneracy is maximal, reflecting the large number of independent antiferromagnetic configurations on the disconnected loops. This classification accounts for all observed phase diagrams of the Fibonacci graph, examples of which are shown in Fig.~\ref{fig:phases}.

\subsection{Fermionic hopping model}

\subsubsection{Ground-state correlations}

To probe the robustness of the fermionic ground state to deviations from the special system sizes $N=2^l$, we examine density-density correlations, which provide a direct measure of how changes in the underlying graph geometry are reflected in the ground-state structure. For negative $s$, the density-density correlations exhibit the familiar short-range behaviour, decaying rapidly with distance around the ring geometry. As $s$ increases towards the long-range regime, pronounced peaks emerge at the graph-connected distances, reflecting the growing influence of the sparse long-range hopping network. For sufficiently positive $s$, the correlations become strongly concentrated on these graph-connected distances, mirroring the underlying graph structure. Representative results for the power-of-two graph are shown in Appendix~\ref{app:ferm-robust}.

The fermionic hopping model exhibits the same geometric sensitivity to system size as the TFIM, since the long-range hopping structure is determined by the same interaction graph. Consequently, the behaviour is again controlled by $\gcd(N,d)$, with the dominant interaction distance setting the effective geometry. For $s<0$, nearest-neighbour hopping dominates and the system remains largely insensitive to variations in $N$. For $s>0$, however, changes in $\gcd(N,d_f)$ can qualitatively alter the effective geometry, leading to dramatic changes in the density-density correlations and low-energy spectrum. In particular, system sizes with $\gcd(N,d_f)=1$ retain a ring-like geometry, while $\gcd(N,d_f)>1$ leads to fragmentation into disconnected loops. Representative examples illustrating this sensitivity are presented in Appendix~\ref{app:ferm-robust}.

\section{Conclusion \& Outlook}


In this work, we investigated the low-energy many-body phenomena in a broad class of sparse graphs. While the power-of-$p$ family preserves a self-similar structure, varying $p$ introduces qualitatively distinct physics. In particular, the power-of-three graph highlights the role of geometric frustration, which stabilises the antiferromagnetic state across the entire phase diagram. All other even $p$ and odd $p$ fall into the two qualitative classes represented by $p = 2$ and $p = 3$, respectively. This identifies the power-of-two graph as the most structurally rich example within this class. The recursive structure of the power-of-$p$ family also makes these graphs naturally suited to experimental realization in neutral-atom shuttling architectures \cite{Bluvstein_2023,Kuriyattil_2025}.

The Fibonacci graph breaks the discrete self-similarity of the power-of-$p$ family, making it more difficult to realize in neutral-atom shuttling architectures, where the efficiency of the protocol relies on this structure. Nevertheless, the Fibonacci graph remains naturally accessible in cavity-QED platforms with direct control over the inclusion and removal of individual bonds \cite{Bentsen_2019_sparse}. This absence of self-similarity gives rise to an exact mirroring between the short- and long-range regimes. Unlike the power-of-$p$ graphs, where tuning across $s=0$ changes the effective geometry from a ring-like structure to disconnected loops, a Fibonacci reordering maps the graph onto itself for all $s$, reversing only the hierarchy of interaction distances. Consequently, the low-energy physics for positive and negative $s$ is directly mirrored, yielding symmetric phase diagrams and identical 1D quantum Ising transitions in both limits.

Overall, our results demonstrate that our geometry-based perspective provides a framework for understanding the phases and criticality of sparse long-range systems, and establish the power-of-two graph as a particularly rich and experimentally relevant case within this broader class. An important future direction is to determine how far this effective geometry framework extends beyond equilibrium, and whether the geometries identified here continue to govern quantum quenches, and information spreading across phase boundaries. Equally intriguing is the question of integrability in these sparse models: understanding their exactly solvable limits, the onset of integrability breaking as $s$ is tuned, and possible regimes of near-integrable behaviour. 

Finally, while holographic ideas have already been explored in non-$p$-adic spin systems, such as Ising models on hyperbolic lattices \cite{Asaduzzaman_2022,Breuckmann_2020,Okunishi_2024} and tensor-network constructions \cite{Jahn2022tensornetworkmodels,jahn2018holographycriticalitymatchgatetensor,Sahay_2025}, the connection between power-of-$p$ models and $p$-adic field theories \cite{Gubser_2017,heydeman2017tensornetworkspadicfields} suggests that these spin systems provide a promising setting for future studies of $p$-adic holography.\\
In compliance with EPSRC's open access initiative, the data in this paper will be available from \cite{data_open_access}.
\begin{acknowledgments}
 We thank Gregory Bentsen, Chris Hooley, Monika Schleier-Smith, and Toonyawat Angkhanawin for helpful discussions. This work was supported by the EPSRC through the QQQS programme grant (EP/Y01510X/1) and by the QCi3 hub (EP/Z53318X/1).
 \end{acknowledgments}

\onecolumngrid
\clearpage


\setcounter{section}{0}
\setcounter{equation}{0}
\setcounter{figure}{0}
\setcounter{table}{0}
\setcounter{page}{1}
\renewcommand{\theequation}{S\arabic{equation}}
\renewcommand{\thefigure}{S\arabic{figure}}
\renewcommand{\thesection}{S\arabic{section}}

\appendix
\section{Classical phase diagram extension to power-of-$p$ graphs}\label{app:classical}

\subsection{Even $p$}\label{app:even_p_gaps}
All of the scaling and criticality arguments presented in \cite{gunning_2025} for the power-of-two classical phase diagram carry over verbatim to the power-of-$p$ family upon the replacement $2\to p$. Importantly, the furthest-neighbour bonds
 form closed loops of length \(p\). In the main text we clarified that the ground-state for $s\to0^-$ and $s \geq0$ will be the recursive ground-state generated from effective blocks of size $p$. We now discuss the first excited state above this ground-state for $s<0$ and $s>0$ and provide analytical proofs that this energy gap will collapse to gapless regions at all points except $s=0$.
 
The first critical point which occurs at $s=-2$ for the power-of-two graph, following the analytical form,
\begin{equation}
\Delta E_\infty
=
J\left(
1-\frac{2^{s+1}}{1-2^{s+1}}
\right),
\end{equation}
can be extended to the general even $p$ case. The classical Hamiltonian can be written as,
\begin{equation}
H = \frac{J}{4}\sum_{i=1}^{N}\sum_{d\in \mathscr{D}_p} d^s\,\sigma_i \sigma_{i+d},
\qquad \sigma_i=\pm 1.
\end{equation}

We again compare the antiferromagnetic reference state $\sigma_i=(-1)^i$ with the two-domain-wall excitation obtained by shifting the pattern by one site on a contiguous block. The energy difference is
\begin{equation}
\Delta E = E[\tau]-E[\sigma]
= \frac{J}{4}\sum_{d\in\mathscr{D}_p} d^s \sum_{i=1}^{N}
\left(\tau_i\tau_{i+d}-\sigma_i\sigma_{i+d}\right).
\end{equation}

For a shell of fixed distance $d>1$, only the bonds crossing the two domain walls contribute. Each wall contributes $d$ such bonds, so the total number of affected bonds is $2d$. Each affected bond changes from $+1$ to $-1$, giving an energy shift of $-J/2$ per bond. Therefore the contribution of shell $d$ is
\begin{equation}
\Delta E_d = -\frac{J}{2}(2d)d^s = -J\,d^{s+1}.
\end{equation}

Specialising to the power-of-$p$ shells $d=p^m$, we obtain
\begin{equation}
\Delta E_{p^m} = -J\,p^{m(s+1)}.
\end{equation}

Summing over all shells then gives the finite-size gap
\begin{equation}
\Delta E_N^{(p)}
=
J\left[
1-\sum_{m=1}^{L-1} p^{m(s+1)}
\right].
\end{equation}
Equivalently,
\begin{equation}
\Delta E_N^{(p)}
=
J\left[
1-\frac{p^{s+1}\left(1-p^{(L-1)(s+1)}\right)}{1-p^{s+1}}
\right].
\end{equation}

In the thermodynamic limit, provided $s<-1$ so that the geometric series converges, this becomes
\begin{equation}
\Delta E_\infty^{(p)}
=
J\left(
1-\frac{p^{s+1}}{1-p^{s+1}}
\right).
\end{equation}

The gap therefore closes when
\begin{equation}
\frac{p^{s+1}}{1-p^{s+1}}=1,
\end{equation}
which gives
\begin{equation}
p^{s+1}=\frac{1}{2}
\qquad \Longrightarrow \qquad
s_c(p)=-1-\log_p 2.
\end{equation}

For $s > s_c(p)$, a single-spin flip in the patterned ground state breaks a small number of dominant bonds adjacent to that site in the $p$-loop. The number of such bonds depends on $p$: for $p=2$ a single bond is broken, while for all $p>2$ two bonds are broken. The leading contribution to the excitation energy therefore scales as,
\begin{equation}
\Delta E_{s<0}(N) \sim k\, \frac{J}{2} \left(\frac{N}{p}\right)^s,
\end{equation}
where $k=1$ for $p=2$ and $k=2$ for all $p>2$.

 Conversely, on the $s>0$ side the lowest-energy excitation is obtained by flipping all $p$ spins on a furthest-neighbour loop; such a $p$-spin block flip avoids breaking the $d_{\max}$ bonds and instead frustrates $O(p)$ nearest-neighbor bonds, producing
\[
\Delta E_{s>0}(N)\ \sim\ p\,J\Big(\frac{p}{N}\Big)^{s},
\]
 Equating the relevant competing energy scales yields the same logarithmic finite-size scaling for the pseudocritical exponent,
\[
s_c(N)\propto \frac{1}{\log(N/p)} \xrightarrow[N\to\infty]{} 0,
\]
so the thermodynamic critical point remains at \(s=0\) for all finite \(p\).

\begin{figure}
    \centering
    \includegraphics[width=0.5\linewidth]{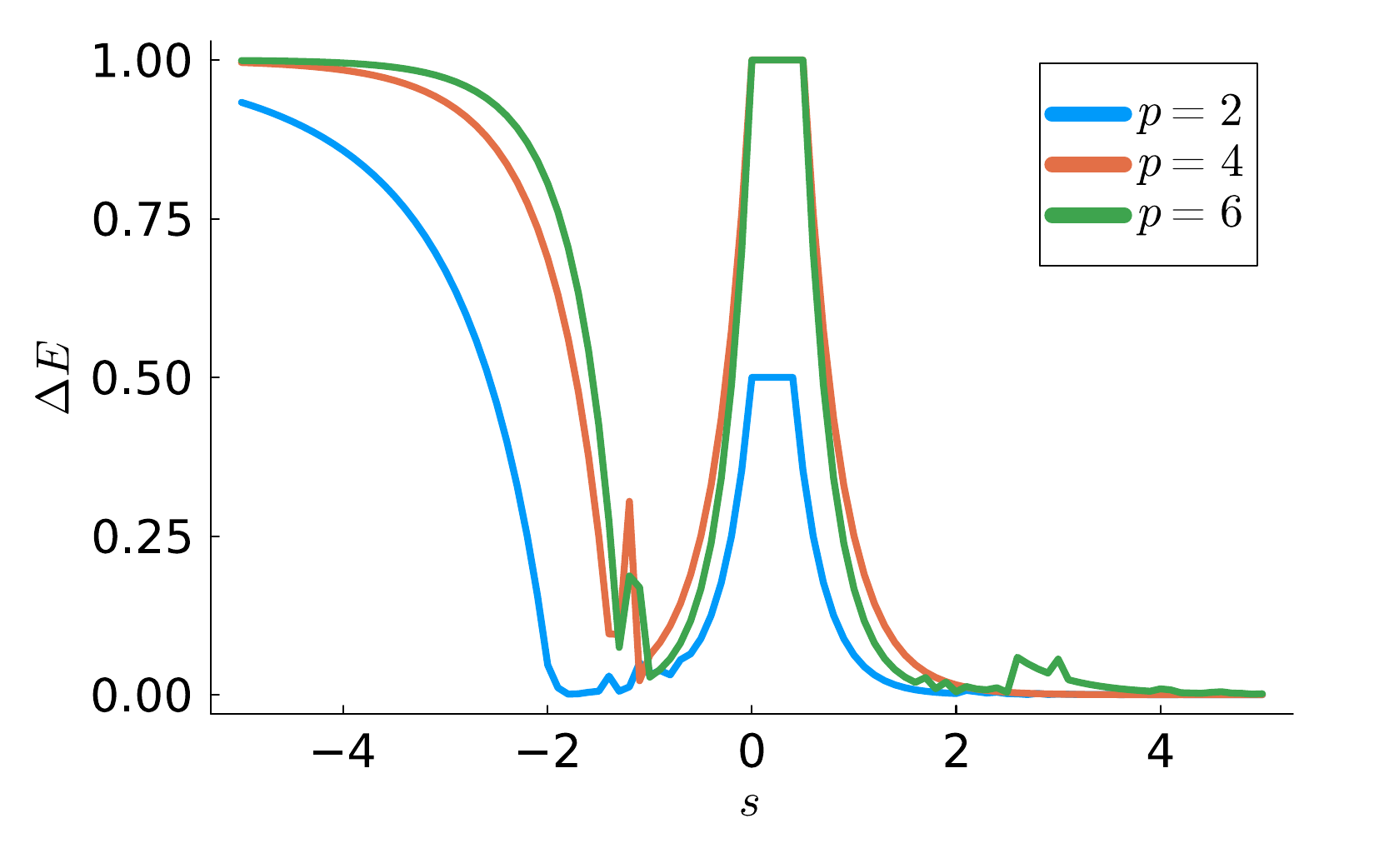}
    \caption{Numerically computed energy gaps for $p=2, N=32$, $p=4,N=64$, $p=6,N=216$ using the Monte Carlo Markov Chain method, which shows good
agreement with the analytical results of Fig.~\ref{fig:phases}(a), except for numerical errors and finite-size effects that arise in the
simulation in the gapless regimes}
    \label{fig:app-mcmc-p}
\end{figure}

Numerical agreement with these analytical predictions is shown in Fig.~\ref{fig:app-mcmc-p}, where the cases $p=2,4,6$ follow the expected behaviour and exhibit the ground-states characterised above.

\subsection{Odd $p$}\label{app:odd_p_gaps}

Here we show that all odd-$p$ power-of-$p$ graphs share the same qualitative classical phase diagram as the $p=3$ case. We consider our classical Ising Hamiltonian Eq.~\eqref{eq:H_IM} for $B=0$.

For any allowed distance $d=p^m$, the bonds $(i,i+d)$ decompose the ring into
\begin{equation}
\gcd(N,d)=d
\end{equation}
disconnected cycles, each of length
\begin{equation}
\ell_d=\frac{N}{d}=p^{L-m},
\end{equation}
which is odd since $p$ is odd. On an odd cycle, an antiferromagnetic arrangement cannot satisfy all bonds, and at least one bond must be frustrated. Thus each shell must contain at least $d$ frustrated bonds in total.

Now consider the nearest-neighbour AFM configuration on the odd ring, i.e. the alternating pattern with one domain wall. Since every allowed distance $d$ is odd, one has $\sigma_{i+d}=-\sigma_i$ everywhere except when the bond crosses the domain wall. In each of the $d$ cycles generated by the distance-$d$ shell, exactly one bond crosses this mismatch, so the AFM frustrates exactly one bond per cycle. It therefore saturates the minimum possible frustration for every shell individually.

Since the full Hamiltonian is a positive weighted sum over these shells, the AFM minimizes each shell separately and hence remains a ground state for all $s$. This proves that odd-$p$ graphs do not undergo a ground-state transition as $s$ is tuned.

For $s>0$, the dominant shell is the furthest-neighbour one, $d_f=N/p$. This decomposes the system into $N/p$ odd loops of length $p$. States that minimize these dominant loops but differ in the subleading shells become asymptotically degenerate with the AFM. With the positive-$s$ normalization, the first correction is suppressed as $p^{-s}$, so the excitation gap collapses as
\begin{equation}
\Delta E \sim p^{-s}.
\end{equation}
Thus all odd-$p$ graphs share the same qualitative phase structure: the AFM remains a ground state for all $s$, while the $s>0$ regime becomes gapless in the limit $s\to+\infty$.

\subsection{Monte Carlo Results for Fibonacci Phase Diagrams}

As the model is classical, we can simulate much larger system sizes than are accessible to exact diagonalization by using Markov-chain Monte Carlo \cite{krauth_2006,robert_MCMC_2011}. 

\paragraph{Metropolis Algorithm.} The algorithm works by setting an initial random configuration of up and down spins. An attempt is made to flip each spin in the chain sequentially, if a flip is energetically favorable it is accepted and if a flip is unfavorable it is rejected unless a randomly generated number $c \in [0,1]$ is less than the Boltzmann weighting $e^{-\beta \Delta E}$ associated with the spin flip. As we are interested in the low lying spectrum (ground-state and first excited state), we set the temperature low, or correspondingly set $\beta$ high. 

\begin{figure}[h!]
    \centering
    \includegraphics[width=0.7\linewidth]{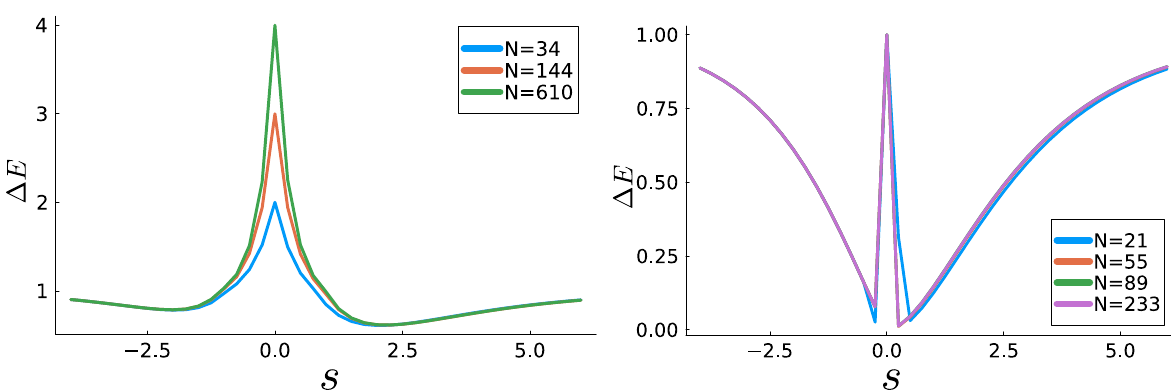}
  \caption{Numerically computed energy gaps obtained using the Monte Carlo Markov Chain method, showing good agreement with the analytical phase diagrams of Fig.~\ref{fig:phases}. The left panel corresponds to odd Fibonacci system sizes, while the right panel corresponds to even Fibonacci system sizes.}
    \label{fig:fib-mcmc}
\end{figure}

In Fig.~\ref{fig:fib-mcmc}, there is excellent agreement between the numerical results and the analytical results of Fig.~\ref{fig:phases}. Our analytical method below is the optimal method to consistently scale to the thermodynamic limit.

\section{Numerical results for quantum phase diagrams}\label{app:quantum}

\subsection{Even $p$}\label{app:even_p_quantum}

\begin{figure}[h!]
    \centering
    \includegraphics[width=0.8\linewidth]{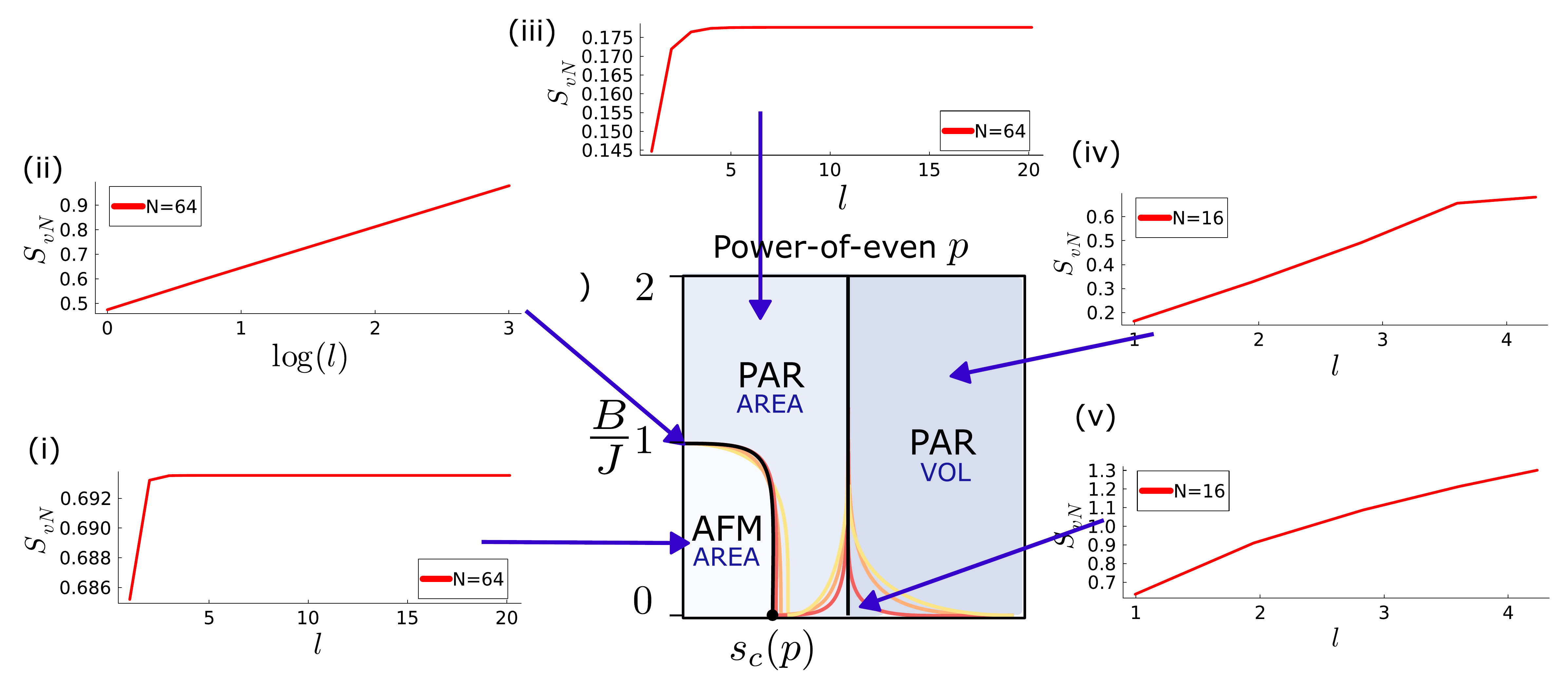}
    \caption{\textit{Quantum phase diagram for power-of-four graph.}
Entanglement scaling for the $p=4$ case is shown, demonstrating agreement with the general even $p$ phase structure. Five distinct regions are identified: (i) area-law scaling in the AFM phase, (ii) logarithmic scaling at the 2D Ising critical point, (iii) area-law scaling in the paramagnetic phase for $s<0$, (iv) intermediate finite-size scaling, and (v) volume-law scaling in the paramagnetic phase for $s>0$.}
    \label{fig:app-pwr4}
\end{figure}

Numerical corroboration for the $p=2$ graph can be found in the Supplemental Material of \cite{gunning_2025}. For $p=4$, system sizes scale as $N=16,64,\dots$, such that only one system size is accessible in all regimes to both exact diagonalization (ED) and density matrix renormalization group (DMRG) methods. For $N=64$ we only find convergence in the solvable limits. In Fig.~\ref{fig:app-pwr4}, we show numerical agreement with the general structure of Fig.~\ref{fig:quantumphases}. 

The entanglement scaling across five regions is in full agreement with the $p=2$ phase diagram: (i) area-law scaling in the AFM phase, (ii) logarithmic scaling at the 2D Ising critical point, (iii) area law-scaling in the PM phase at $s<0$, (iv) intermediate scaling due to finite-size effects, and (v) volume-law scaling in the PM phase at $s>0$.

\subsection{Odd $p$}\label{app:odd_p_quantum}

We use density matrix renormalization group (DMRG) and matrix product states (MPS) to access ground-state properties for system sizes beyond the reach of exact diagonalization (e.g. $N=27,81$). Results are corroborated with our SKQD method for the regime $B\ll1$.

\begin{figure}[h!]
    \centering
    \includegraphics[width=0.9\linewidth]{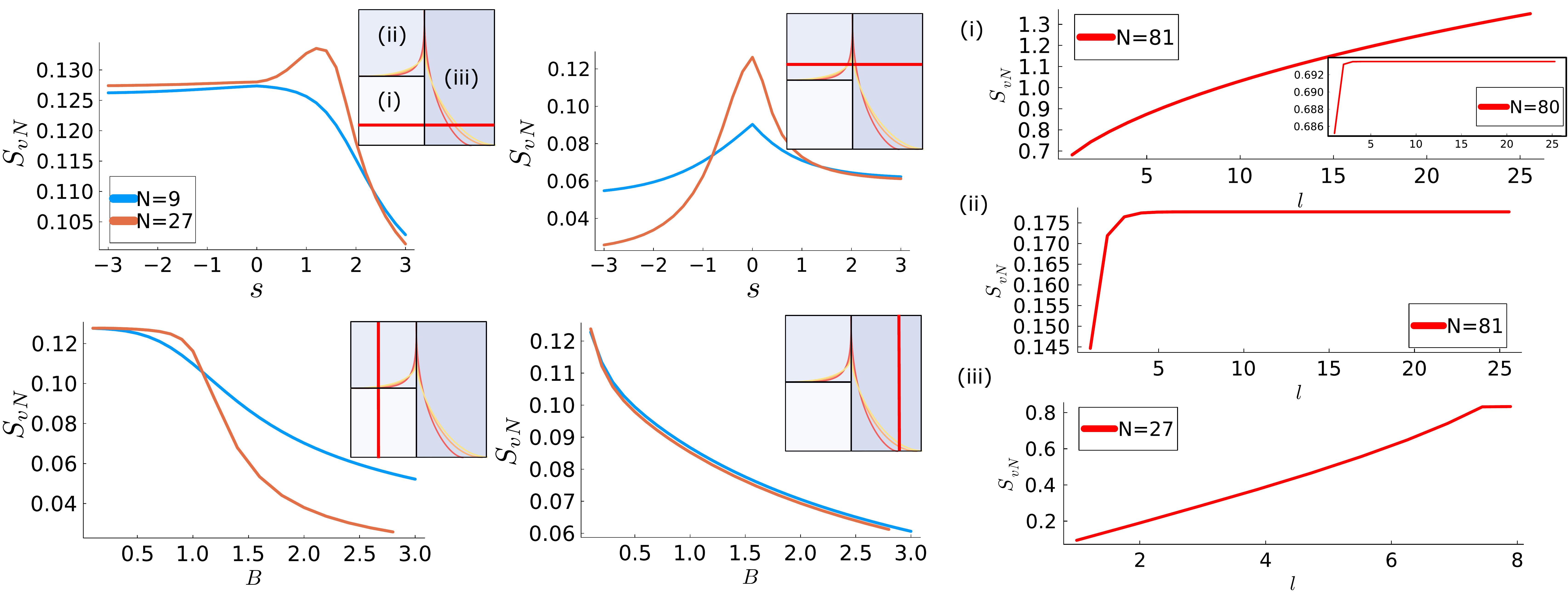}
    \caption{ Half-chain entanglement entropy $S_{vN}$ as a function of coupling parameter $s$ and transverse field strength $B$
capturing the phase structure of the quantum power-of-$3$ model. Four representative slices are shown at $B/J = 0.2, 1.5$ (top
row) and $s = -5, 5$ (bottom row), capturing distinct behaviors across the diagram. These slices correspond to key regimes highlighted in the schematic phase diagram of Fig.~\ref{fig:quantumphases}(b), including regions of area-law and volume-law scaling, as well as the distinct ground-state structure (AFM and PM).}
    \label{fig:app-svn-pwr3}
\end{figure}

A schematic of the quantum phase diagram for $p=3$ is shown in Fig.~\ref{fig:quantumphases}(b) and corroborated in Fig.~\ref{fig:app-svn-pwr3}. In panel (a), we consider the regime of weak quantum fluctuations ($B \ll 1$), where the behaviour closely reflects the classical phase diagram: for $s<0$ the system is in the AFM phase, while for $s>0$ the classical gapless region allows the transverse field to readily induce excitations, resulting in a paramagnetic phase for all finite $B$.

In panel (b), we explore the strong-fluctuation regime, where the ground state is paramagnetic for all $s$. Here, the entanglement scaling distinguishes the phases: for $s<0$ the system exhibits area-law behaviour, while for $s>0$ it exhibits volume-law scaling in the original site ordering. Panel (c) shows the quantum phase transition at $B=J$, which lies in the 2D Ising universality class for sufficiently negative $s$. In panel (d), no phase transition is observed for finite $B$ when $s>0$.

Regions (i) and (ii) both exhibit area-law scaling, as shown in the figure. However, since only odd system sizes are available, this behaviour is not cleanly resolved at accessible sizes. As discussed in Sec.~\ref{sec:robustness}, small variations in system size do not qualitatively affect the phase diagram. We therefore include results for both $N=81$ and $N=80$, where the even system size clearly exhibits area-law behaviour. Region (iii) shows volume-law scaling, as indicated in the figure.

\subsection{Fibonacci graphs}

\subsubsection{Odd $N$}\label{app:odd_fib_quantum}
\begin{figure}[h!]
    \centering
    \includegraphics[width=0.8\linewidth]{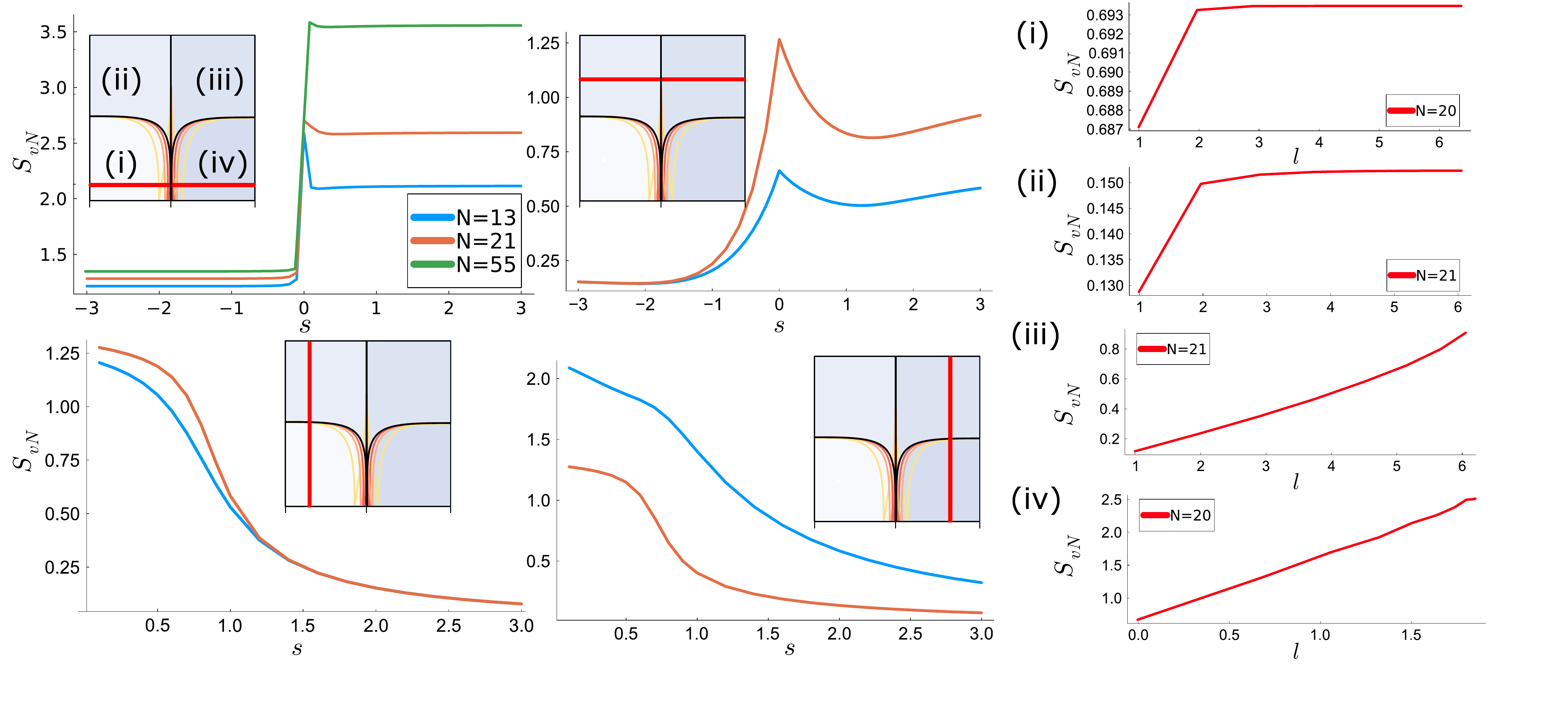}
    \caption{Half-chain entanglement entropy $S_{vN}$ as a function of coupling parameter $s$ and transverse field strength $B$
capturing the phase structure of the quantum Fibonacci model for $N$ odd. Four representative slices are shown at $B/J = 0.1, 1.5$ (top
row) and $s = -5, 5$ (bottom row), capturing distinct behaviors across the diagram. These slices correspond to key regimes highlighted in the schematic phase diagram of Fig.~\ref{fig:quantumphases}(d), including regions of area-law and volume-law scaling, as well as the distinct ground-state structure (AFM, Fibonacci AFM and PM).}
    \label{fig:app-phases-fib-odd}
\end{figure}

Numerical corroboration for the Fibonacci graphs is shown in Fig.~\ref{fig:app-phases-fib-odd} for odd system sizes $N_{\rm fib}=13, 21, 55$. Panel (a) shows the regime of weak quantum fluctuations, where SKQD methods allow access to system sizes up to $N=55$, beyond the reach of reliable MPS convergence. In this regime, a single phase transition is observed at $s=0$, from an area-law AFM phase (i) to a volume-law Fibonacci AFM phase (iv).

Panel (b) shows the strong-fluctuation regime, which again exhibits a single transition at $s=0$, from an area-law paramagnetic phase (ii) to a volume-law paramagnetic phase (iii). Panel (c) shows the transition from AFM to paramagnet at $B=J$, which belongs to the 2D Ising universality class for sufficiently negative $s$. Remarkably, panel (d) reveals an analogous transition from the Fibonacci AFM to the paramagnet at $B=J$, also in the 2D Ising universality class, despite the underlying model appearing long-range and higher dimensional.

\subsubsection{Even $N$}\label{app:even_fib_quantum}

\begin{figure}[h!]
    \centering
    \includegraphics[width=0.85\linewidth]{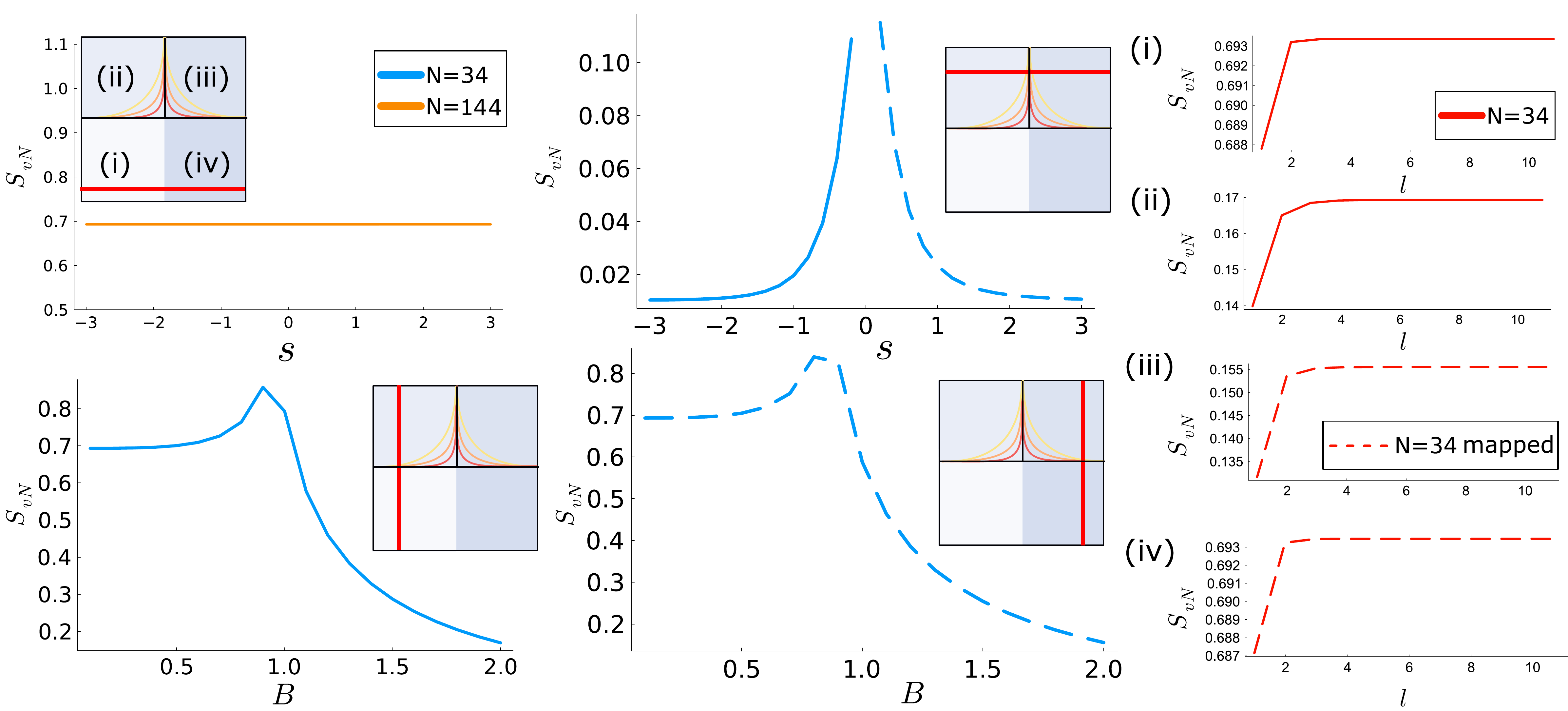}
    \caption{Half-chain entanglement entropy $S_{vN}$ as a function of coupling parameter $s$ and transverse field strength $B$
capturing the phase structure of the quantum Fibonacci model for $N$ even. Four representative slices are shown at $B/J = 0.1, 8.0$ (top
row) and $s = -5, 5$ (bottom row), capturing distinct behaviors across the diagram. These slices correspond to key regimes highlighted in the schematic phase diagram of Fig.~\ref{fig:quantumphases}(c), including regions of area-law and volume-law scaling, as well as the distinct ground-state structure (AFM and PM). Dashed lines indicate Fibonacci mapped results.}
    \label{fig:app-phases-fib-even}
\end{figure}

Numerical corroboration for the Fibonacci graphs with even system sizes is shown in Fig.~\ref{fig:app-phases-fib-even} for $N_{\rm fib}=34,144$. This case provides a particularly clear example of the role of effective geometry. For $s<0$, the dominant interactions are nearest-neighbour and the geometry is the original one-dimensional ring. For $s>0$, we exploit the Fibonacci reordering, which maps sites that are furthest neighbours in the original ordering to adjacent sites in the reordered chain.

Importantly, the Fibonacci reordering is not just an approximation valid in the $s\to+\infty$ limit. It maps the full Fibonacci graph back onto itself. To see this, take $N=F_n$ and let the largest allowed distance be $d_f=F_{n-2}$. Since $\gcd(F_n,F_{n-2})=1$, repeatedly stepping by $d_f$ visits every site exactly once. We can therefore use this sequence to define a new ordering of the sites. This is unlike the power-of-$p$ graphs where the monna map holds exactly only for $s \to \infty$.

In this reordered chain, bonds that were at distance $d_f$ in the original labelling become nearest-neighbour bonds. The remaining allowed Fibonacci distances are also mapped to allowed Fibonacci distances, but in the opposite order of strength. Thus the $s>0$ Hamiltonian is exactly the same Fibonacci graph written in a different site ordering, with the interaction hierarchy reversed relative to the $s<0$ limit. This exact graph equivalence explains the symmetry of the even-$N$ Fibonacci phase diagram about $s=0$.

The reordering is also essential for numerical convergence in the $s>0$ regime. In the Fibonacci-ordered basis, the entanglement scaling is area-law, as expected for a gapped one-dimensional effective geometry. However, this should be distinguished from the scaling with respect to the original site ordering. In the original ordering, a half-chain cut intersects an extensive number of strong Fibonacci bonds, $O(N)$, rather than $O(1)$. The same state may therefore appear volume-law in the original ordering, even though it is area-law in the effective Fibonacci geometry. We account for this distinction when assigning the entanglement scaling below.

Panel (a) shows the regime of weak quantum fluctuations, where SKQD methods allow access to system sizes up to $N=144$, beyond the reach of reliable MPS convergence. In this regime, no phase transition occurs, reflecting the classical phase diagram in which the AFM remains the ground state for all $s$. The change across $s=0$ is therefore not a change of ordering, but a change in the effective geometry: the system goes from an area-law AFM phase in the original ring geometry (i) to a Fibonacci AFM phase which is area-law in the reordered geometry but volume-law with respect to the original ordering (iv).

Panel (b) shows the strong-fluctuation regime. Here the system is paramagnetic on both sides of $s=0$, but the entanglement structure changes from an area-law paramagnet for $s<0$ (ii) to a volume-law paramagnet for $s>0$ (iii), when measured in the original site ordering. Panel (c) shows the transition from the AFM to the paramagnet at $B=J$, which lies in the 2D Ising universality class for sufficiently negative $s$. Panel (d) reveals the analogous transition from the AFM to the paramagnet at $B=J$, also in the 2D Ising universality class. This is a striking result: although the $s>0$ model appears long-ranged in the original ordering, its effective geometry is again one-dimensional, making the shared universality class natural from the geometry-first perspective.

\section{Fermionic Hopping Model}

\subsection{PWR3 Energy gap}\label{app-ferm-pwr3}
\begin{figure}[h!]
    \centering
    \includegraphics[width=0.5\linewidth]{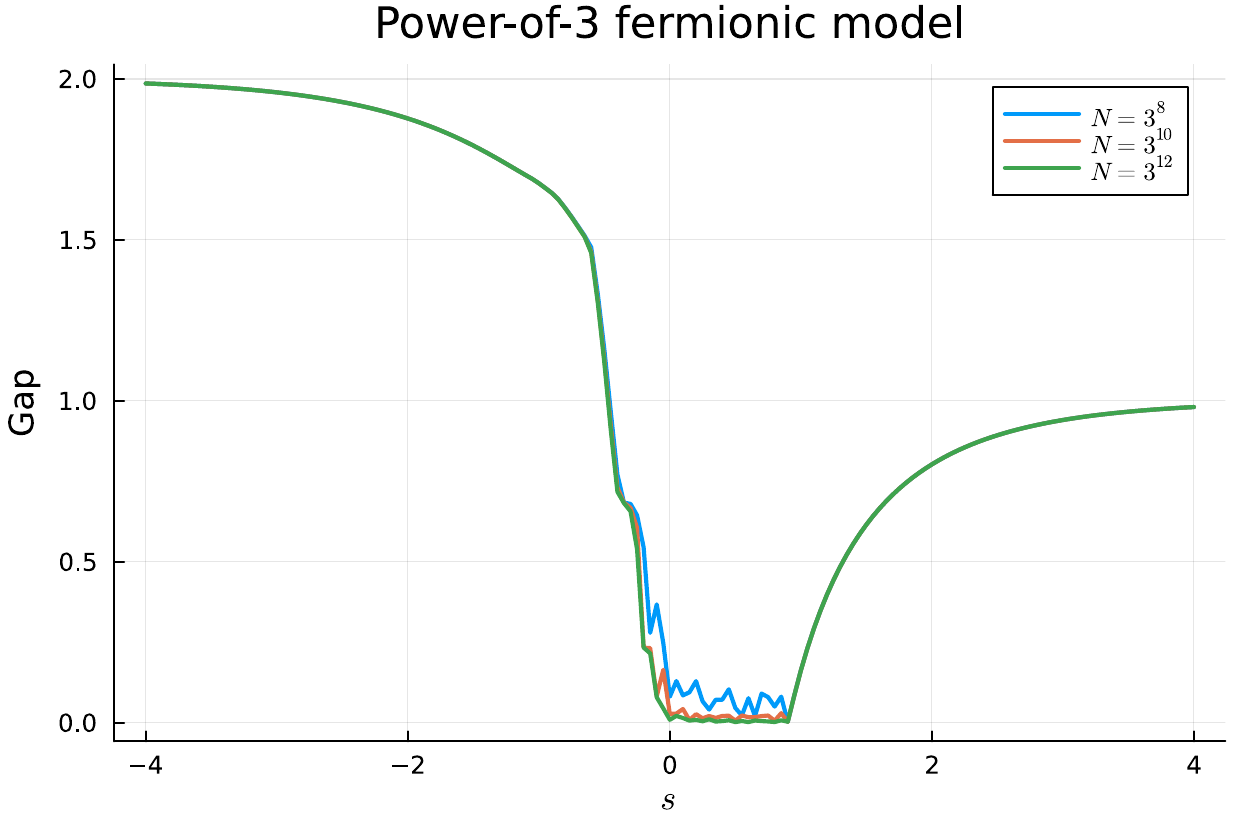}
    \caption{Spectral gap of the fermionic power-of-three (PWR3) model as a function of interaction exponent $s$ for increasing system sizes $N$. For $s<0$, the gap remains finite and converged with system size. For $s>0$, the gap scales toward zero over an extended interval of positive $s$, indicating a gapless thermodynamic regime, before reopening beyond a critical value $s_c$.}
    \label{fig:app-ferm}
\end{figure}

For $s<0$, the spectral gap remains finite and well converged with increasing system size, consistent with the analytic prediction that the quasiparticle spectrum remains gapped. By contrast, for $s>0$ the minimum gap decreases systematically with system size over an extended range of positive $s$, indicating a gapless thermodynamic regime despite the absence of an exact analytic mode closing. Beyond a critical value $s_c\approx 0.9$, the gap reopens and the system becomes gapped once again. Numerical evidence for this behaviour is shown in Fig.~\ref{fig:app-ferm}.

\subsection{Sensitivity away from special system sizes for fermionic hopping model}\label{app:ferm-robust}

\subsubsection{Ground-state correlations}

\begin{figure}[h!]
    \centering
    \includegraphics[width=0.6\linewidth]{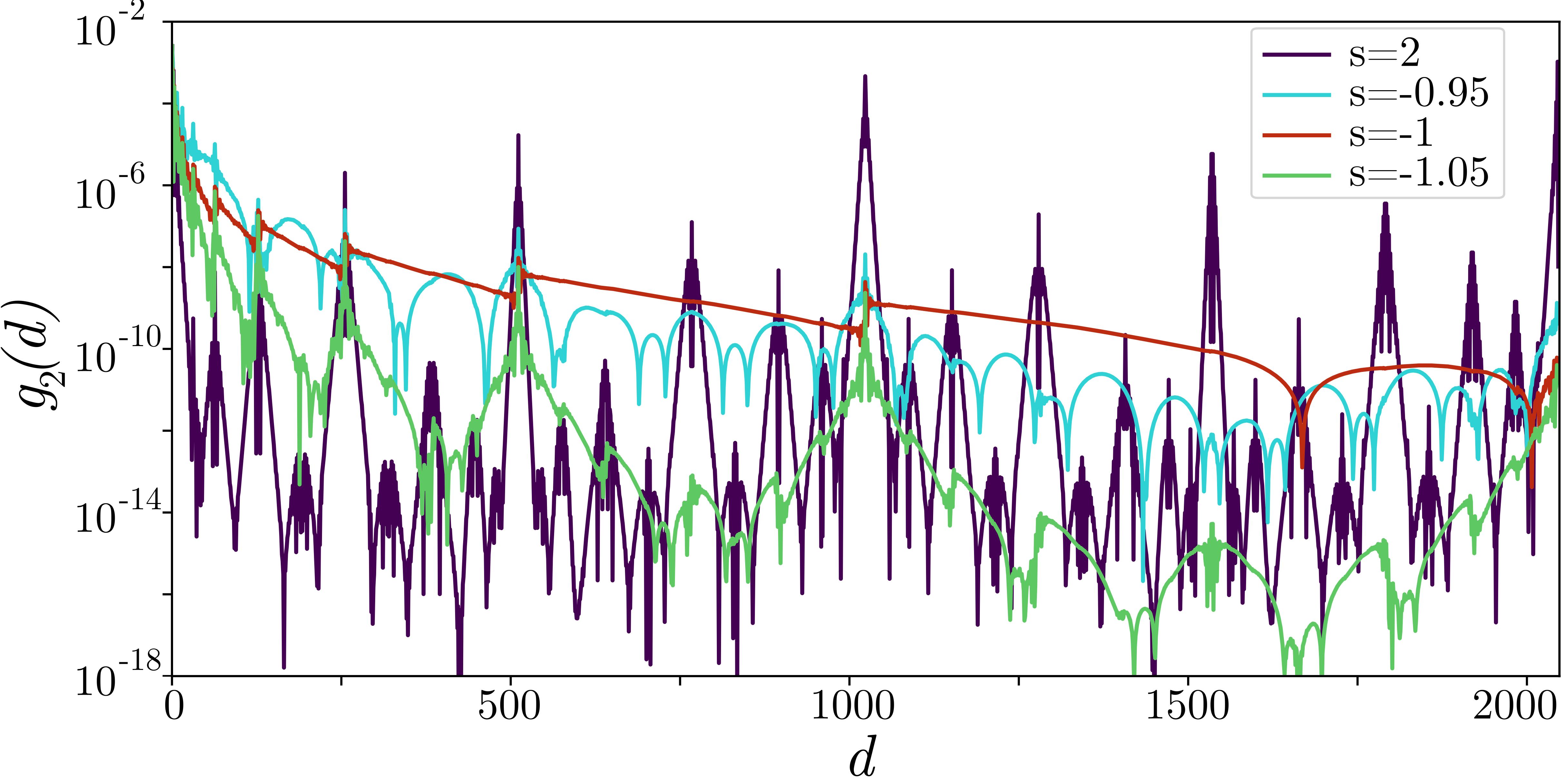}
  \caption{Density-density correlations Eq.~\eqref{eq:densitydensity} for the fermionic power-of-two graph \eqref{eq:H_F} with pairing strength $\Delta=1$ and system size $N=2^{12}=4096$. The correlations evolve from exponentially decaying ring-like behaviour in the short-range regime to algebraic decay at the gap closing $s=-1$. As $s$ approaches $0$, pronounced peaks develop at graph-connected distances while the overall decay weakens. For sufficiently positive $s$, correlations become strongly concentrated on the graph-connected distances, reflecting the dominance of the sparse long-range connectivity. The qualitative behaviour is unchanged for larger system sizes up to $N\approx2^{19}$.}
    \label{fig:pwr2_corrs}
\end{figure}

To assess the robustness of the fermionic ground state to variations in system size away from the special values $N=2^q$, we analyse the density-density correlation function. As a ground-state observable, it provides a direct probe of the effective geometry induced by the hopping graph and allows us to determine whether the correlation structure remains stable under small changes in $N$. The correlations we investigate are,
\begin{equation}
    \begin{aligned}
        g_2\left(r, r+d \right) = g_2\left(d\right) &= \langle n_r n_{r+d} \rangle - \langle n_r \rangle \langle n_{r+d} \rangle \\
        &= \lvert \langle c^{\dagger}_r c^{\dagger}_{r+d} \rangle \rvert^2 - \lvert \langle c^{\dagger}_r c_{r+d} \rangle \rvert^2.
    \end{aligned}
    \label{eq:densitydensity}
\end{equation}
Density-density correlations are connected to the $C_{r,r+d}^{zz}$ correlation functions of the corresponding Ising model through the Jordan-Wigner transformation \cite{vodola2016long}. Representative results for the power-of-two graph at $\Delta=1$ are shown in Fig.~\ref{fig:pwr2_corrs}. In the short-range regime, correlations exhibit the expected one-dimensional behaviour, decaying exponentially with distance. As the system approaches the gap closing at $s=-1$, this decay is known to become algebraic, reflecting the onset of criticality. Across all values of $s$, the sparse interaction graph leaves a clear signature in the form of pronounced peaks at the graph-connected distances $d=2^n$. These peaks are superimposed on a smooth correlation background inherited from the nearest-neighbour pairing term. For increasing positive $s$, the graph-connected peaks become progressively more dominant, while the underlying background weakens, reflecting the growing importance of the sparse long-range connectivity. In the limit $s\rightarrow\infty$, the correlations become concentrated on the dominant long-range bonds, consistent with the decomposition of the graph into disconnected antipodal pairs.

\subsubsection{Sensitivity to variations away from special system sizes}

The geometric mechanism discussed in Sec.~\ref{sec:robustness} carries over directly to the fermionic hopping model. Since the hopping graph is determined by the same set of interaction distances, the effective geometry in the long-range regime is again controlled by $\gcd(N,d_f)$, where $d_f$ denotes the dominant hopping distance.

For the power-of-two graph with system sizes $N=2^q$, the dominant hopping distance is $d_f=2^{q-1}$, yielding $\gcd(N,d_f)=N/2$ when $s>0$. The hopping graph therefore decomposes into $N/2$ disconnected dimers in the $s\gg0$ limit. By contrast, increasing the system size by a single site to $N=2^q+1$ restores $\gcd(N,d_f)=1$, causing the dominant long-range bonds to form a single connected loop.

\begin{figure}[h!]
    \centering
    \includegraphics[width=0.6\linewidth]{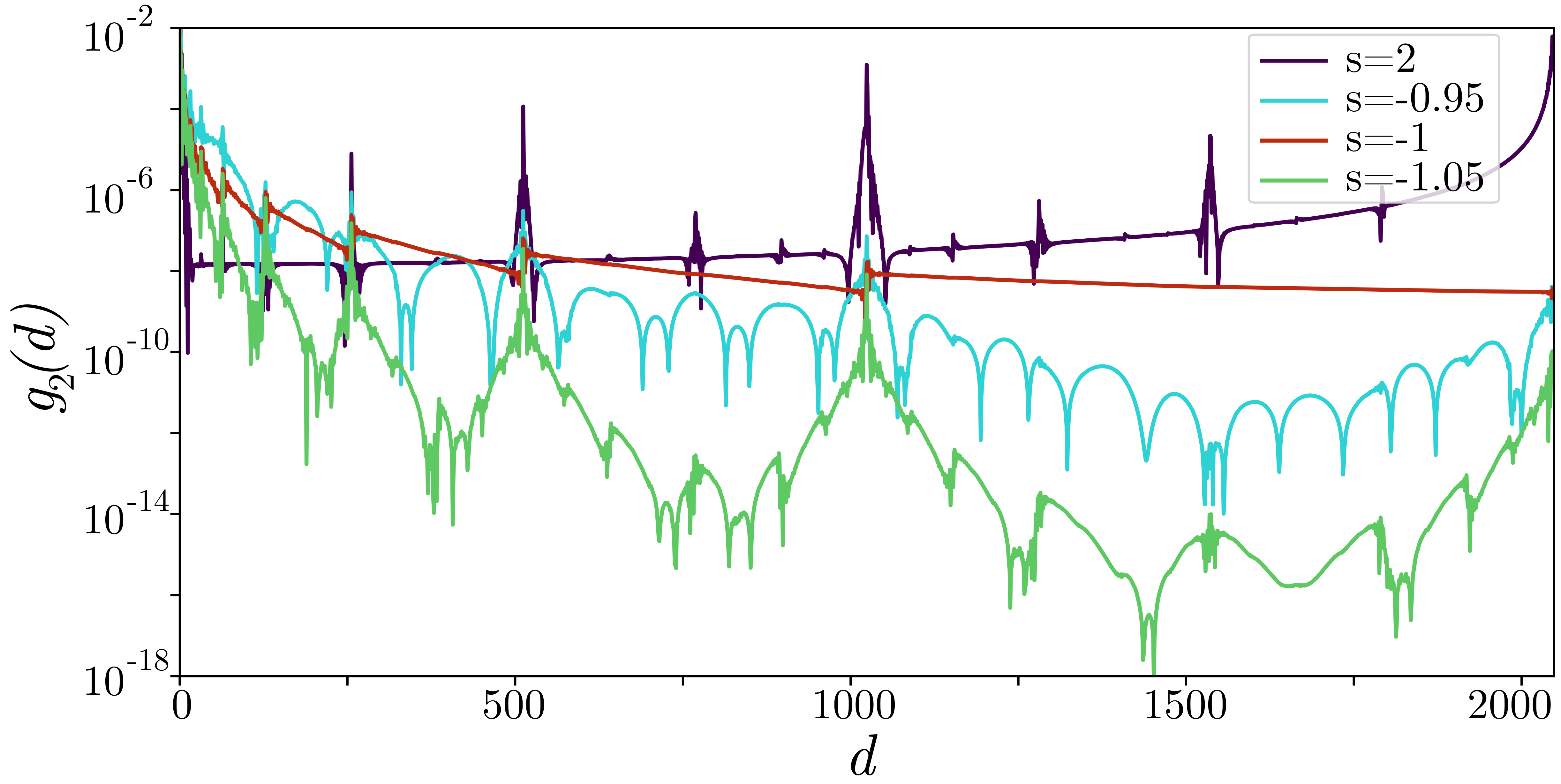}
    \caption{
Density-density correlations Eq.~\eqref{eq:densitydensity} for the fermionic power-of-two graph with $\Delta=1$ and $N=2^{12}+1=4097$. Compared with Fig.~\ref{fig:pwr2_corrs}, the addition of a single site changes the long-range effective geometry from disconnected antipodal pairs to a single connected loop. While the short-range regime remains largely unchanged, the long-range correlations are qualitatively modified, illustrating the strong system-size sensitivity of the sparse graph geometry.
}
    \label{fig:pwr2_corrs_syssize}
\end{figure}

The resulting change in the density-density correlations is shown in Fig.~\ref{fig:pwr2_corrs_syssize}. In the short-range regime ($s<0$), where nearest-neighbour hopping dominates, the correlations are largely unaffected by the change in system size. In the long-range regime ($s>0$), however, the difference is dramatic. Whereas the $N=2^q$ system exhibits correlations concentrated on disconnected antipodal pairs, the $N=2^q+1$ system develops a ring-like correlation structure inherited from the connected long-range loop. Thus, despite the change of only a single site, the correlation pattern is qualitatively modified even in the thermodynamic limit.

This sensitivity closely parallels the edge-state phenomena discussed in Ref.~\cite{vodola2016long}, where small modifications to the effective geometry produce macroscopic changes in the low-energy physics.

\bibliography{bibliography}
\end{document}